\documentclass[lettersize,journal]{IEEEtran}

\usepackage[caption=false,font=normalsize,labelfont=sf,textfont=sf]{subfig}
\usepackage{mathrsfs}
\usepackage{latexsym}
\usepackage{graphicx}
\usepackage{epsfig}
\usepackage{array}
\usepackage{amsmath}
\usepackage{amssymb}
\usepackage{xcolor}
\usepackage{amsthm}
\usepackage{enumerate}

\usepackage{algorithm}
\usepackage{algorithmic}
\usepackage{booktabs}
\usepackage{cite}
\usepackage{bm}
\usepackage{balance}
\usepackage{verbatim}
\usepackage{epstopdf}
\usepackage{setspace}
\usepackage{multicol}
\usepackage{amsmath,lipsum}
\usepackage{cuted}
\usepackage{multirow}
\usepackage{makecell}

\graphicspath{{figures/}} 

\usepackage{url}

\begin{document}

	\title{Two-Stage Refinement Sparse Channel Estimation for Reconfigurable Intelligent Metasurface Antenna (RIMSA) Massive MIMO}
	
	\author{ Yakun Ma, Hui-Ming Wang,~\IEEEmembership{Senior Member,~IEEE},  Jiaping He, and Qingli Yan
	\thanks{Yakun Ma, Hui-Ming Wang and Jiaping He are with the School of Information and Communications Engineering, Xi'an Jiaotong University,
	Xi'an 710049, China (e-mail: mayakun1998@163.com; xjbswhm@gmail.com; jiapinghe2021@163.com).
	
	Qingli Yan is with the School of Computer Science \& Technology, Xi’an University of Posts \& Telecommunications, Xi'an 710121, China (e-mail: yql@xupt.edu.cn).
		}
	}
\maketitle

\begin{abstract}
To meet the increasing demands for high data rates and large capacity, next generation wireless communication systems require transceivers equipped with a large number of antennas. Massive multiple-input multiple-output (MIMO) with metasurface antennas has emerged as a promising solution. In this paper, we investigate the channel estimation problem for the emerging reconfigurable intelligent metasurface antenna (RIMSA) array systems. Specifically, we develop a two-stage refinement (TSR) channel estimation method based on the compressed sensing (CS) principle. In the first stage, we exploit the antenna structure of RIMSA to receive pilots by setting identical phase response vectors across all RIMSAs. In this manner, the coherence of the measurement matrix  under the CS framework is reduced and the channel estimation performance is improved.
However, this special design introduces channel direction-of-arrival (DoA) estimation ambiguity and yields an ambiguous candidate DoA set. In the second stage, we optimize the phase responses of the metamaterial elements to resolve the ambiguity and accurately estimate the DoAs and channel coefficients.
Overall, the estimation accuracy is improved in the first stage at the cost  of ambiguity, and  this ambiguity is eliminated in the second stage. We illustrate the performance advantages of the TSR method by presenting the numerical results and comparing it with the existing methods.
\end{abstract}


\section{Introduction}
With the rapid development of mobile communications, users are demanding higher data rates and capacity. Key technologies such as multiple-input multiple-output (MIMO) are being deployed in the fifth-generation (5G) mobile communications network to support various application scenarios such as enhanced broadband transmission and high-reliability and low-latency communication \cite{1}. As an extension of MIMO technology, extra-large massive MIMO (XL-MIMO)  has been introduced to play a significant role in the development of the sixth-generation (6G) mobile communications \cite{2}. However, XL-MIMO systems tend to become complex, costly, and energy-consuming \cite{3}. Therefore, energy-efficient MIMO technologies are required to make the vision of high power efficiency, lower latency, higher reliability and wider coverage a reality.

In this context, reconfigurable intelligent surfaces (RISs) have attracted increasing attention. Specifically, a RIS consists of a large number of densely packed subwavelength passive electromagnetic (EM) metamaterial elements \cite{4}, whose physical parameters can be adjusted independently to tailor the propagation characteristics of EM waves (EMWs). The  RIS enables flexible manipulation of EM-wave propagation and radiation patterns by regulating the resonant characteristics of electronic components (e.g., adjustable reactive  and resistive elements) in each metamaterial element \cite{5,6,7}. Owing to its excellent EMW manipulation ability, a RIS is able to reconfigure or regulate the wireless propagation environment~\cite{5}.  

\begin{figure}[!t]
	\centering
	\subfloat[]{\includegraphics[width=0.32\textwidth]{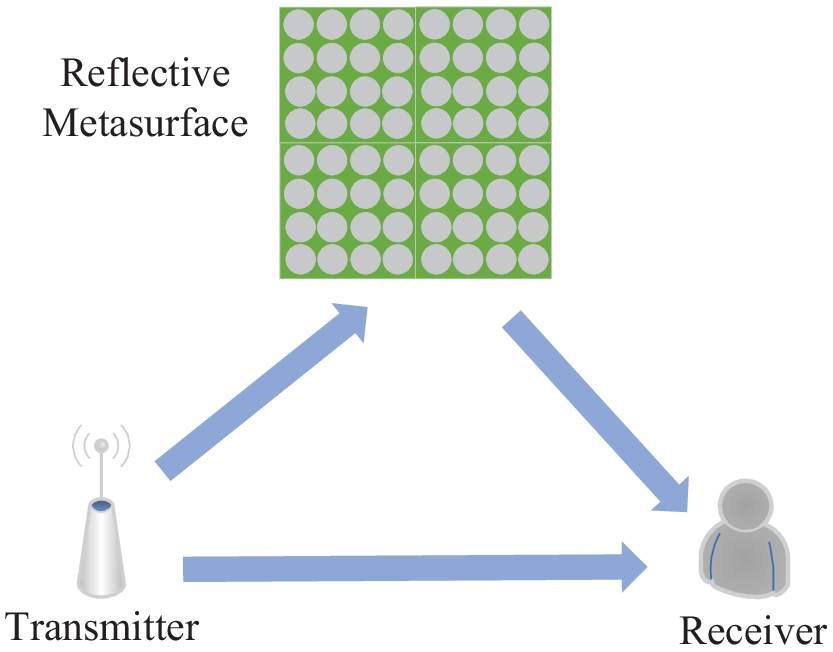} 	}\\
	\subfloat[]{\includegraphics[width=0.32\textwidth]{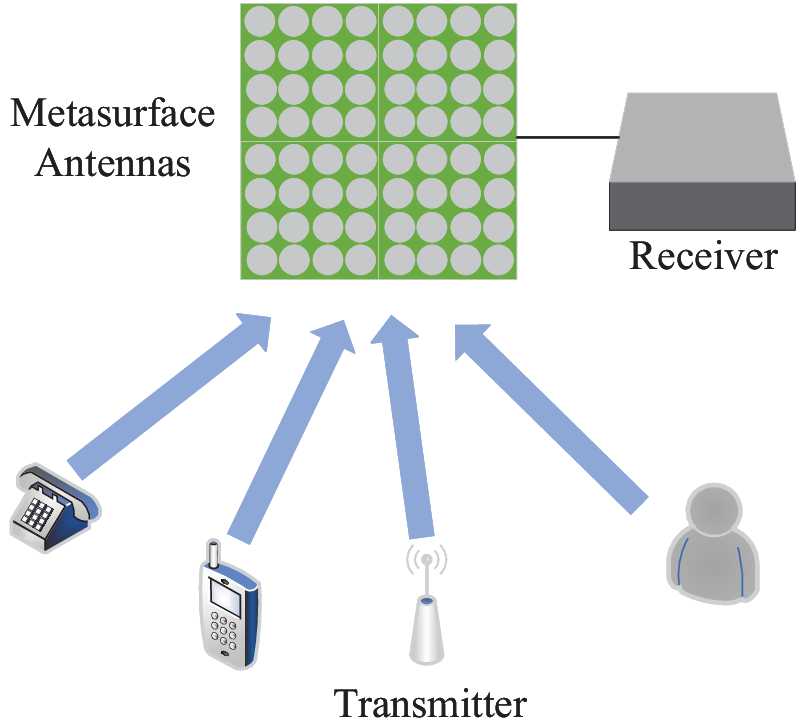}  }
	\caption{(a) Reflective metasurface; (b) Reconfigurable metasurface used as a receive antenna.}
	\label{fig_1}
\end{figure}


Currently, the wide-ranging applications of RIS are centered on its role as a passive reflective surface, which is deployed between a transmitter and a receiver, as illustrated in Fig. \ref{fig_1}(a). A typical RIS adjusts the phase of an EM signal incident on the surface and reflects it back into the propagation environment, thereby helping the signals between the transmitter and receiver to overcome harsh propagation conditions and expanding network coverage \cite{8}. Recent studies have shown that  as a passive device, RIS performs well in increasing communication rates \cite{9,10,11} and enhancing energy efficiency \cite{12,13}, or even improving physical-layer security \cite{121,122}. However, in the case of RIS-aided communication links, the double fading effect due to reflection makes the reflected signal at the receiver undergo significant attenuation \cite{15}. To address this issue, active RIS architectures have been proposed in \cite{123,14,16} to enhance reflection power. Nevertheless, active RIS suffers from the same problems of high cost and low energy efficiency due to active components.

The application of metasurfaces goes beyond the traditional way as a reflector. Antennas constructed using RIS can directly control the phase and amplitude of radiated EMWs, enabling signal transmission and reception in a real-time and dynamically configurable manner. As shown in Fig. \ref{fig_1}(b), metasurface antennas are deployed at a receiver. By adjusting the antenna response, the metasurface antenna can process the impinging signals through analog beamforming at the antenna front end without  the need for additional hardware. 
Two typical architectures of such metasurface antennas are dynamic metasurface antennas (DMAs) \cite{19, 20} and reconfigurable holographic surfaces (RHSs) \cite{18}. However, both architectures are commonly based on series-fed structures, where the metamaterial elements are excited sequentially. Specifically, the signal propagation along the feeding waveguide in DMAs introduces additional frequency selectivity, while poor two-dimensional (2D) port isolation makes direct 2D implementations challenging. 
In RHSs, the feed-generated reference wave propagates along the metasurface as in a leaky-wave antenna, so the phase depends on the element position, while only the amplitude can be directly adjusted.

\begin{figure}[!t]
	\centering
	\includegraphics[width=3.0 in]{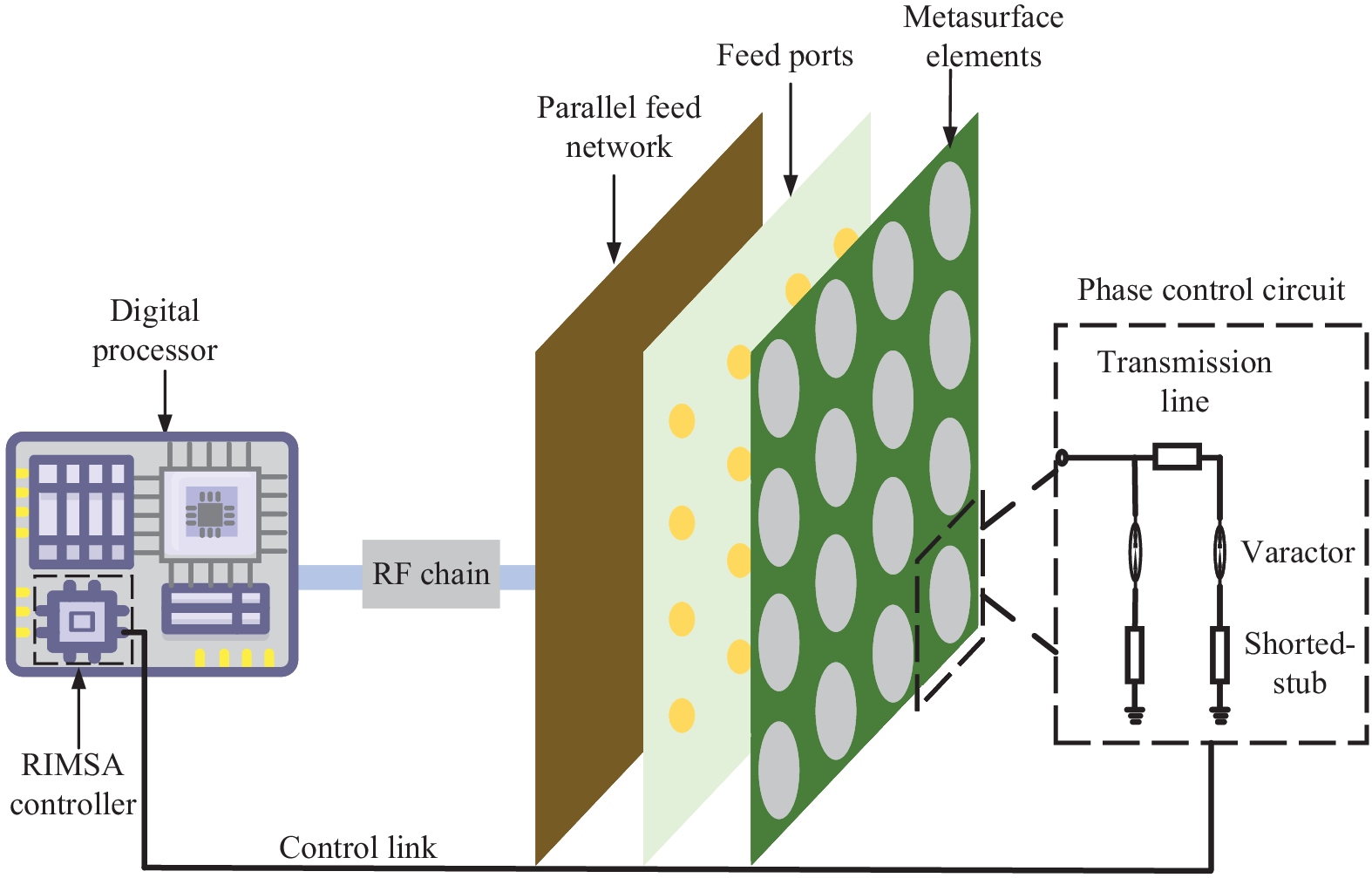}
	\caption{The architecture of RIMSA.}
	\label{fig_2}
\end{figure}

Recently, a reconfigurable intelligent metasurface antenna (RIMSA) architecture has been proposed in \cite{WangZhang2026, HuangWang2026, add4, add3}. Unlike DMA and RHS, the RIMSA employs a bottom-coaxial-feed architecture with a parallel feed network, where each metamaterial element is directly excited by a bottom feed, as shown in Fig.~\ref{fig_2}. This architecture ensures better isolation between ports and facilitates the realization of 2D arrays. Furthermore, by exciting all elements simultaneously via a coaxial feed network, RIMSA mitigates the frequency selectivity of the antenna response. The reconfiguration of each element is regulated by applying a direct current (DC) voltage to a varactor diode, realizing continuous phase responses. Since each RIMSA antenna connects multiple elements to a single radio frequency (RF) chain, the receiver structure naturally resembles a hybrid analog-digital MIMO system \cite{52}.

Acquiring accurate channel state information (CSI) is indispensable for fully realizing the potential of metasurface-enabled wireless systems, and developing efficient channel estimation schemes remains an active research topic. 
For RIS-assisted communications, channel estimation is particularly challenging due to the passive nature of the reflecting elements. To address this, the authors in \cite{HamidrezaKhaleghi2025} proposed an optimized separate channel estimation strategy leveraging deterministic channel models and optimal RIS configurations to significantly lower the pilot overhead. From a compressed sensing (CS) perspective, \cite{DavidWilliam2024} introduced a matching pursuit with phase rotation algorithm for RIS-aided millimeter wave systems, which efficiently estimates the steering vectors and complex channel gains by deploying a few active elements at the RIS panel.

Beyond passive reflectors, metasurface-based antenna arrays have drawn considerable attention. The work in \cite{29} proposes a least-squares estimator for a planar holographic MIMO (HMIMO) array. 
Considering the non-ideal nature of practical hardware, \cite{AnzhengTang2026} explicitly investigated HMIMO channel estimation under severe mutual coupling effects, reformulating the task as a vector factorization problem and developing a hybrid message passing algorithm. 
For DMA-assisted systems, the channel estimation problem is investigated in \cite{35} under the minimum-mean-square-error criterion. To enhance the estimation performance, the DMA phase-shifting matrix and the pilot signals are optimized accordingly.
Furthermore, \cite{RuoyuZhang2025} proposed  a tensor-based channel estimation framework for DMA-assisted orthogonal frequency division multiplexing systems that leverages angular sparsity. 
Specifically for the direction-of-arrival (DoA) estimation of metasurface antennas, 
the radiation beams generated by a HMIMO array are exploited to estimate channel parameters under a line-of-sight-dominated channel model \cite{add1}.
In addition, a MUSIC-based algorithm is employed in \cite{add2} to estimate the DoA information.

Furthermore, given the structural similarity between RIMSA arrays and hybrid MIMO systems, channel estimation methods for hybrid MIMO serve as a baseline. Prior works \cite{39,40, JavadMirzaei2021,41,42, BiqingQi2019,42a,43,44} have addressed CS-based channel estimation for hybrid MIMO under sparse multipath channels. Specifically, non-uniformly quantized angle grids \cite{39}, convex optimization \cite{40}, and hybrid analog-digital beamforming designs under the minimum mean squared error criterion \cite{JavadMirzaei2021} have been employed for measurement matrix design to reduce coherence and enhance estimation performance. To mitigate grid mismatch, grid refinement approaches, such as the distributed grid matching pursuit (DGMP) \cite{41}, Newton refinement methods \cite{42}, and iterative off-grid error compensation algorithms \cite{BiqingQi2019}, have been proposed. Regarding gridless approaches, iterative reweighted (IR) methods \cite{43} and multi-resolution codebooks \cite{44} have been developed. 
However, these existing methods typically focus on reconstructing the channel over a full-dimension angular dictionary. Consequently, the mutual coherence of the measurement matrix is inherently constrained by the large size of the dictionary, limiting further improvements in estimation accuracy.

In this paper, we propose a CS-based two-stage refinement (TSR) channel estimation method tailored for the RIMSA-equipped receiver.
Different from prior RIS-aided methods \cite{HamidrezaKhaleghi2025, DavidWilliam2024} that estimate cascaded reflected channels and from DMA/HMIMO estimators \cite{29, AnzhengTang2026, 35, RuoyuZhang2025, add1, add2} tailored to series-fed or continuous aperture metasurface architectures, this work focuses on the parallel-fed RIMSA receiver. More importantly, unlike conventional CS-based hybrid MIMO channel estimation methods \cite{39,40, JavadMirzaei2021,41,42, BiqingQi2019,42a,43,44} that optimize the measurement matrix or refine grids over a fixed full-dimensional dictionary, the proposed TSR method leverages the specific periodicity of the RIMSA layout to directly reduce the column dimension of the measurement matrix in the first stage, thereby alleviating the mutual coherence bottleneck of full-dictionary CS methods.
Specifically, by configuring identical phase-shifting vectors across all RIMSAs, the response vectors associated with different physical DoAs are intentionally folded into a reduced set of principal phase responses. This design decreases the mutual coherence of the effective measurement matrix and improves the reliability of first-stage DoA estimation.
In addition, we optimize the phase-shifting vectors to ensure that the spatial response of the RIMSA is sufficiently flat across the entire DoA range of interest to guarantee robust received power---a practical constraint often overlooked in previous works. Although the first stage introduces DoA ambiguity and yields a candidate angle set, it effectively provides a high-precision initial path subspace. 
Subsequently, in the second stage, the phase response of each metamaterial element is further optimized to resolve this ambiguity and enable accurate estimation of the DoAs and channel coefficients from the candidate set.
Consequently, the proposed method achieves improved channel estimation accuracy compared with conventional CS-based channel estimation methods.
The contributions of this paper are as follows: 
\begin{itemize}
	\item[$\bullet$] 
	We propose a CS-based TSR channel estimation framework, which decomposes the estimation process into subspace acquisition and ambiguity resolution. This method alleviates the mutual coherence limitation of full-dimension dictionary methods, enabling high-precision estimation with limited RF chains.
	
	\item[$\bullet$] 
	In the first stage, the structural periodicity of the RIMSA arrays is exploited to compress the measurement matrix. By configuring identical phase-shifting vectors across antennas, we significantly reduce the lower bound of mutual coherence, enabling high-precision acquisition of the signal subspace. Additionally, the RIMSA spatial response is optimized to ensure flat coverage, effectively preventing signal loss caused by random pattern nulls.
	
	\item[$\bullet$]
	In the second stage, a refinement mechanism is developed to resolve the DoA ambiguity introduced by the specific phase configuration in the first stage. 
	A Riemannian manifold optimization algorithm is utilized to optimize the pilot measurement matrix, thereby minimizing the mutual coherence and improving the accuracy of DoA and channel coefficient estimation from the candidate set.
	
	\item[$\bullet$]
	We present comprehensive numerical simulation results and analysis of the proposed method. Additionally, a comparison with existing channel estimation methods is provided. The numerical results demonstrate that the proposed method outperforms existing methods and achieves superior estimation accuracy for RIMSA-based receiver systems.
\end{itemize}

The remainder of this paper is organized as follows. In Section II, we introduce the RIMSA array model in detail. The channel estimation problem based on the antenna structure is introduced in Section III. In Section IV, we present the proposed TSR method. Then, we analyze the computational complexity and feasibility in Section V. Simulation results and analyses are given in Section VI. Section VII concludes the paper.

\emph{Notations:}
In this paper, scalars are denoted by italic letters (e.g., $a$), vectors by boldface lowercase letters (e.g., $\mathbf{a}$), and matrices by boldface uppercase letters (e.g., $\mathbf{A}$). 
The superscripts $(\cdot)^T$, $(\cdot)^*$, $(\cdot)^H$, and $(\cdot)^{-1}$ denote the transpose, complex conjugate, conjugate transpose, and inverse operators, respectively.
The operator $|\cdot|$ represents the absolute value of a scalar, while $\|\cdot\|_0$, $\|\cdot\|_2$, and $\|\cdot\|_{\mathrm{F}}$ denote the $\ell_0$-norm, $\ell_2$-norm, and Frobenius norm, respectively. The symbol $\odot$ represents the Hadamard product. Furthermore, $\operatorname{diag}(\mathbf{a})$ denotes a diagonal matrix with the elements of vector $\mathbf{a}$ on its main diagonal, and $\operatorname{blkdiag}(\mathbf{A}_1, \dots, \mathbf{A}_N)$ represents a block diagonal matrix formed by matrices $\mathbf{A}_1, \dots, \mathbf{A}_N$. The space $\mathbb{C}^{M \times N}$ refers to the set of $M \times N$ complex-valued matrices. The operator ${\mathrm{vec}}(\cdot)$ denotes the vectorization operator, which stacks the columns of a matrix into a single column vector, and $\mathrm{Re}\{\cdot\}$ denotes the real part of a complex number.
Finally, $\exp(\cdot)$ and $\angle(\cdot)$ denote the element-wise exponential and element-wise phase extraction operators, respectively.

\section{System Model}
In this section, we first describe the RIMSA architecture in Section II-A and then present the corresponding channel and training models in Sections II-B and II-C, respectively.

\subsection{RIMSA Architecture}
	The RIMSA is a flexibly and configurable metasurface antenna, facilitating adaptation and reconfiguration according to the wireless environment. As shown in Fig. \ref{fig_2}, one  RIMSA  primarily comprises metamaterial elements, varactor-based phase-shifting circuits, a parallel feed network, and feed ports. The phase control circuit of each metamaterial element is able to  adjust the phase of EM signals. Specifically, by regulating the DC voltage of the phase control circuits, the capacitance of varactors can be adjusted to change the phase response of the metamaterial element. The phase shift is contingent upon the phase control circuit, which  leads to a continuous phase control of the EMW and analog beamforming. When operating in the receiving mode, the impinging EMW excites all metamaterial elements of a RIMSA simultaneously. The received element-level signals are then combined through the parallel feed network and delivered to the corresponding RF chain for baseband processing. Since the elements are fed in parallel rather than sequentially along a waveguide, the RIMSA avoids the waveguide-induced frequency selectivity typically observed in DMA or RHS architectures.

Conventional hybrid analog-digital MIMO architectures rely on dedicated analog phase shifter networks, which inherently suffer from high power consumption due to the active tuning requirements of phase shifters. In contrast, the radiation/response pattern of a RIMSA is formed by dynamically tunable metamaterial elements. These elements are integrated into a parallel feeding network, which is composed of passive components like varactor diodes and significantly reduces power consumption. Furthermore, due to the limited response speed of phase shifters, traditional hybrid MIMO systems exhibit fixed radiation patterns within a single symbol period. Nevertheless, RIMSA enables sub-symbol-level beam reconfiguration thanks to the nanosecond-level response time of varactor diodes, as described in detail in \cite{add3}. Consequently, the metamaterial elements can be rapidly reconfigured several times by the control circuitry within a single symbol period, facilitating real-time adaptation to dynamic channel conditions. 

\begin{figure}[!t]
	\centering
	\includegraphics[width=0.9\linewidth]{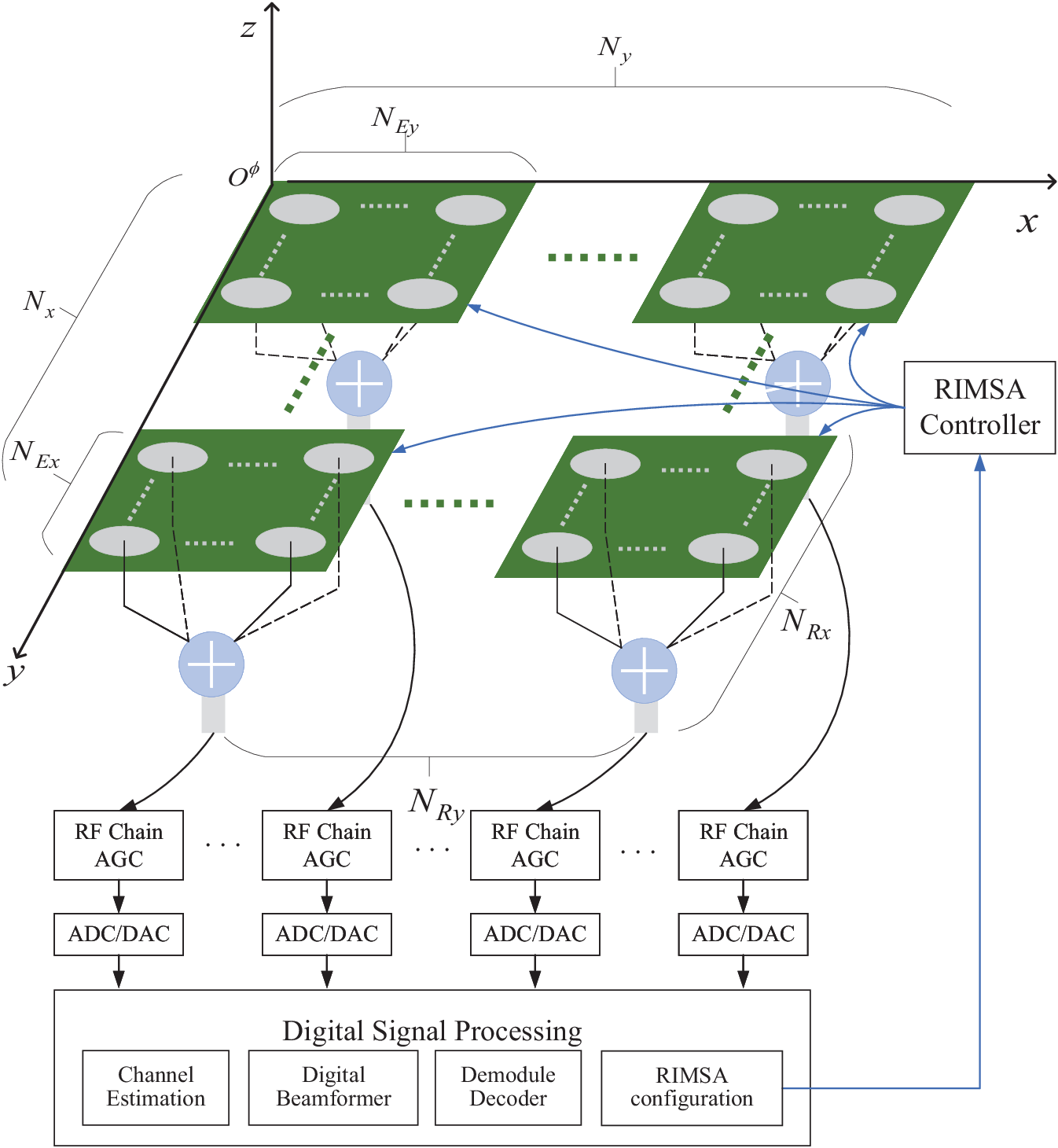}
	\caption{A planar RIMSA array with $N_{R_y}N_{R_x}$ antennas (RF chains) and a total of $N_xN_y$ metamaterial elements, where each RIMSA is also a planar antenna with $N_{Ex}N_{Ey}$ elements.}
	\label{fig_3}
\end{figure}

A general RIMSA array is illustrated in Fig. \ref{fig_3}. Each RIMSA consists of ${N_E} = {N_{Ex}} \times {N_{Ey}}$ densely spaced subwavelength metamaterial elements and is connected to one RF chain. The RIMSA array is composed of ${N_R} = {N_{Rx}} \times {N_{Ry}}$ such metasurface antennas, yielding an inter-antenna spacing of $N_{Ex}d$ along the x-axis and $N_{Ey}d$ along the y-axis, where $d$ is the adjacent element spacing within each RIMSA. Thus, the whole receiver contains $N=N_RN_E$ metamaterial elements and $N_R$ RF chains. Integrated EMW control and reconfiguration is achieved by dynamically configuring the phase responses of the metamaterial elements. Specifically, for ${{n_r} = 1,2,...,{N_R}}$ and ${{n_e} = 1,2,...,{N_E}}$, the response of the ${n_e}$-th metamaterial element in the RIMSA connected to the ${n_r}$-th RF chain is
\begin{equation}
	v_{{n_r},{n_e}} =  e^{ - j\alpha _{{n_r},{n_e}}}, 	\label{deqn_ex1}
\end{equation}
where  ${\alpha _{{n_r},{n_e}}}$ denotes the phase shift determined by the bias voltage applied to the corresponding varactor diode. This response provides an analog weight for the element-level received signal before the signals within the same RIMSA are combined and fed to the associated RF chain. Accordingly, the receive phase-shifting vector of the RIMSA connected to the ${n_r}$-th RF chain is defined as
\begin{equation}
	\label{deqn_ex2}
	{\mathbf{v}}_{{n_r}}^{} = 1/\sqrt {{N_E}}{\left[ {v_{{n_r},1}^{},v_{{n_r},2}^{},...,v_{{n_r},{N_E}}^{}} \right]^T} .
\end{equation}

\subsection{Channel Model}
The propagation channel between the RIMSA array and transmitter is modeled as a narrowband frequency-flat multipath channel with $K$ distinct propagation paths. The parametric channel model ${\mathbf{h}}$ can be expressed as
\begin{equation}
	\label{deqn_ex3}
	{\mathbf{h}} = \sum\limits_{k = 1}^K {{{\mathbf{a}}_k}{\beta _k}},  
\end{equation}
where ${\beta _k}$ denotes the complex channel attenuation, and ${{\mathbf{a}}_k}\in\mathbb{C}^{{N\times 1}}$ represents the array manifold vector corresponding to the 
$k$-th incident path, with $k=1,2,\ldots,K$. The  array manifold vector is structured as
\begin{equation}
	\label{deqn_ex4}
	{{\mathbf{a}}_k} = {\left[ {{\mathbf{a}}_{k,1}^T,\ldots,{\mathbf{a}}_{k,{n_r}}^T,\ldots,{\mathbf{a}}_{k,{N_R}}^T} \right]^T},
\end{equation}
where each ${\mathbf{a}}_{k,{n_r}}\in\mathbb{C}^{N_E\times 1}$, ${n_r} = 1,...,{N_R} $ is the steering vector of the ${n_r}$-th RIMSA as
\begin{equation} 
{\mathbf{a}}_{k,{n_r}}^{} = {\left[ {{e^{ - j2\pi \frac{{{\mathbf{q}}_{{n_r},1}^H{{\mathbf{e}}_k}}}{\lambda }}},...,{e^{ - j2\pi \frac{{{\mathbf{q}}_{{n_r},{N_E}}^H{{\mathbf{e}}_k}}}{\lambda }}}} \right]^T}, 
\end{equation}
where $\lambda$ is the wavelength, ${{\mathbf{e}}_k}$ is the unit vector of $k$-th direction, and ${{\mathbf{q}}_{{n_r},{n_e}}}$ is the position of ${n_e}$-th metamaterial element of ${n_r}$-th RIMSA with
$	{{\mathbf{q}}_{{n_r},{n_e}}} = {\left[ {{d_{x,{n_r},{n_e}}},{d_{y,{n_r},{n_e}}},0} \right]^T}$
where ${{d_{x,{n_r},{n_e}}}}$ and ${{d_{y,{n_r},{n_e}}}}$ represent the $x$-coordinate and $y$-coordinate relative to the array reference point $O$.
The channel model can be written in a more compact matrix form as
\begin{equation}
	\label{deqn_ex6}
	{\mathbf{h}}={\mathbf{A}\bm{\beta}},
\end{equation}
where ${\bm{\beta }} = {\left[ {{\beta _1},...,{\beta _K}} \right]^T}$ is the path fading vector and ${\mathbf{A}} = \left[ {{{\mathbf{a}}_1},...,{{\mathbf{a}}_K}} \right]\in\mathbb{C}^{N\times{K}}$ is the array manifold matrix.

\subsection{Training Model} 
\begin{figure}[!t]
	\centering
	\includegraphics[width=0.95\linewidth]{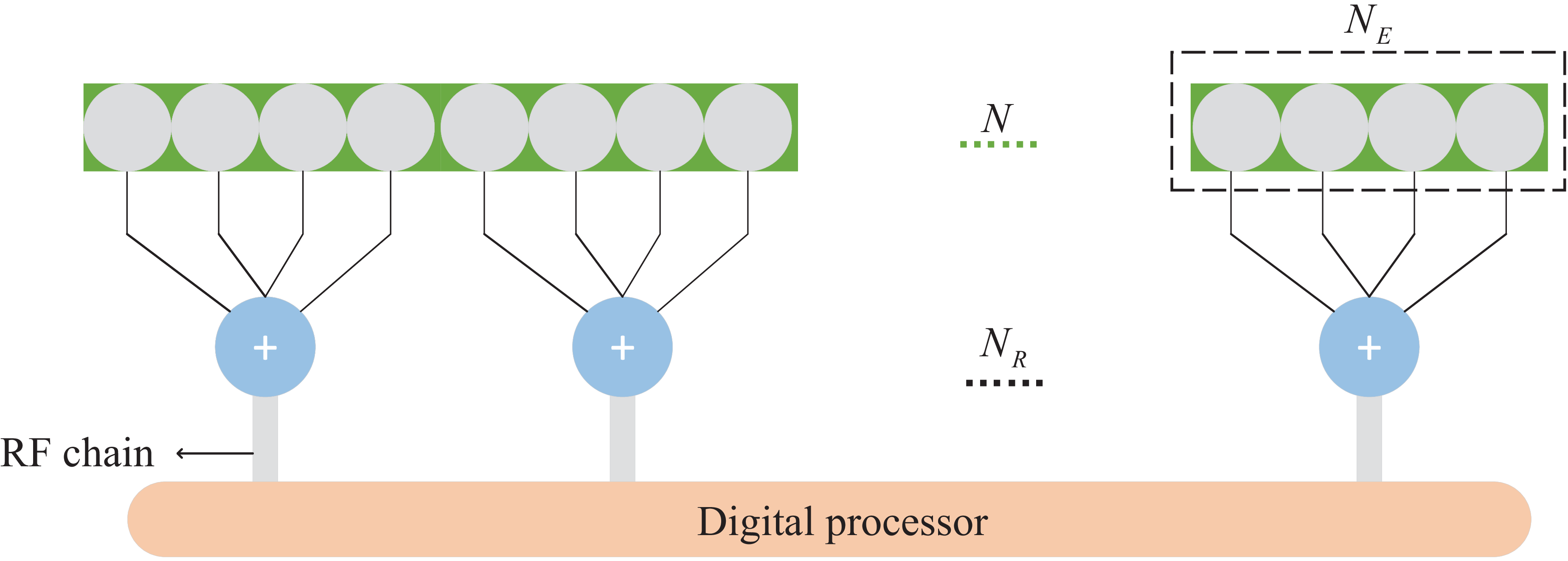}
	\caption{A 1D RIMSA array with $N_R$ antennas and $N_E$ elements in each antenna.}
	\label{fig_4}
\end{figure}

While the 2D planar RIMSA represents the practical hardware architecture, we utilize a 1D RIMSA array model shown in Fig.~\ref{fig_4} to clearly present the mathematical framework of the proposed TSR method without notation clutter.
\footnote{Under ideal conditions, this 1D formulation can be mathematically generalized to a 2D uniform planar array. By leveraging the properties of the Kronecker product, configuring identical phase shifts across 2D RIMSAs decouples the array manifold exactly as in the 1D case, thus achieving the same dimension reduction and ambiguity set size. The ambiguity set size inherently corresponds to the number of metamaterial elements in each RIMSA.}

Consider a far-field transmitter that sends a known deterministic pilot sequence $\mathbf{s} = [s_1,s_2,\ldots,s_L]^T \in \mathbb{C}^{L\times 1}$, where $L$ denotes the pilot length and $s_l$, $l=1,2,\dots,L$, is the pilot symbol transmitted at the $l$-th training instant.
The pilot sequence is normalized as $\frac{1}{L}\sum_{l=1}^{L}|s_l|^2=1.$
Let $P$ denote the transmit power of each pilot symbol. Thus, the actual transmitted pilot symbol at the $l$-th training instant is $\sqrt{P}s_l$.
Let $\Theta  = \left\{ {{\theta _1},...,{\theta _K}} \right\}$ denote the DoA set, where ${\theta _k},k = 1,2,...,K$ represents the DoA of the $k$-th propagation path.
Let ${\mathbf y}_l\in\mathbb{C}^{N_R\times 1}$ denote the received vector at the $l$-th training instant. Then, the received signal can be expressed as
\begin{equation}
	\label{deqn_ex7}
	{\mathbf{y}}_l = \sqrt{P}{\mathbf{V}}^H{\mathbf{h}} s_l + {\mathbf{V}}^H{\mathbf{z}}_l
	= \sqrt{P}{\mathbf{V}}^H{\mathbf{A}}\left( \Theta  \right){\bm{\beta }} s_l + {\hat{\mathbf z}}_l,
\end{equation}
where  ${\mathbf{V}}\in\mathbb{C}^{{N}\times{N_R}}$ represents the phase-shifting matrix of the RIMSA array as
\begin{equation}
	\label{deqn_ex8}
	{\mathbf V} = \left[ {\begin{array}{*{20}{c}}
			{{\mathbf{v}}_1^{}}&{\mathbf{0}_{N_E}}& \ldots &{\mathbf{0}_{N_E}}\\
			{\mathbf{0}_{N_E}}&{{\mathbf{v}}_2}& \ldots &{\mathbf{0}_{N_E}}\\
			\vdots & \vdots & \ddots & \vdots \\
			{\mathbf{0}_{N_E}}&{\mathbf{0}_{N_E}}&{\mathbf{0}_{N_E}}&{{\mathbf{v}}_{N_R}}
	\end{array}} \right],
\end{equation}
with ${\mathbf{v}}_{n_r}\in\mathbb{C}^{{N_E}\times 1}$ in the form of Eq.
(\ref{deqn_ex2}), and each $\mathbf{0}_{N_E}$ denotes a zero vector of dimension $N_E \times 1$.
The term ${\mathbf{z}}_l\in\mathbb{C}^{N\times 1}$ is the additive white Gaussian noise vector at the $l$-th training instant, with ${\hat{\mathbf z}}_l  = {\mathbf{V}}^H{\mathbf{z}}_l$ and ${\mathbf{z}}_l\sim\mathcal{CN}({\mathbf 0},\sigma_z^2{\mathbf I}_{N})$. The noise vectors are assumed to be independent over different training instants. The semi-unitary property ${{\mathbf{V}}^H}{\mathbf{V}} = {{\mathbf{I}}_{N_R}}$ ensures that
${\hat{\mathbf z}}_l\sim\mathcal{CN}({\mathbf 0},\sigma_z^2{\mathbf I}_{N_R})$.
The array manifold matrix is ${\mathbf{A}}\left( \Theta  \right) = \left[ {{\mathbf{a}}\left( {{\theta _1}} \right),...,{\mathbf{a}}\left( {{\theta _K}} \right)} \right]\in\mathbb{C}^{N\times K}$, where ${\mathbf{a}}\left( {{\theta _k}} \right) = {\left[ {1,{e^{ - j2\pi \frac{d}{\lambda }\sin {\theta _k}}},...,{e^{ - j\left( {N - 1} \right)2\pi \frac{d}{\lambda }\sin {\theta _k}}}} \right]^T}$.
By correlating the received signals with the known pilot sequence and normalizing by $\sqrt{P}$, we obtain
\begin{align}
	{\mathbf{\tilde y}}
	= \frac{1}{L\sqrt{P}} \sum\limits_{l = 1}^L {{\mathbf{y}}_l s_l^*}
	= {\mathbf{V}}^H{\mathbf{A}}\left( \Theta  \right){\bm{\beta }}
	+ {\mathbf{\tilde z}},	\label{deqn_ex9}
\end{align}
where the equivalent noise denotes
${\mathbf{\tilde z}} = \frac{1}{L\sqrt{P}}\sum\limits_{l = 1}^L {{\hat{\mathbf z}}_l s_l^*}$ and it follows~${\mathbf{\tilde z}}\sim\mathcal{CN}\left({\mathbf 0},\frac{\sigma_z^2}{LP}{\mathbf I}_{N_R}\right).$

\section{Problem Formulation}
The sparse angular domain channel estimation problem can be reformulated as a joint multipath component separation and complex gain recovery problem. 
Since the array response depends on the direction cosine, let $\psi\in(-\pi/2,\pi/2]$ denote the physical angle and let $\zeta=\sin(\psi)\in(-1,1]$ denote its direction cosine. We uniformly discretize $\zeta$ into $G$ grid points. Specifically, the $g$-th, $g = 1,2,...,G$  direction cosine grid point is $\zeta_g=-1+2g/G$, and the corresponding physical angle is $\psi_g=\arcsin(\zeta_g)$. The resulting overcomplete physical angle grid is defined as
\begin{equation}
	\label{deqn_ex10a}
		{{\Psi }}_G=\left\{\psi_g\mid g=1,2,\ldots,G\right\},
\end{equation}
where $G\gg K$ controls the direction cosine grid resolution. The channel vector ${\mathbf{h}}$ in (\ref{deqn_ex3}) can be approximated \cite{42a} using the extended virtual channel model as
\begin{equation}
	\label{deqn_ex10}
	{\mathbf{h}} \approx {\mathbf{A}}\left( {{{{\Psi }}_G}} \right){{\bm{\beta }}_G},
\end{equation}
where ${{\bm{\beta }}_G} = {\left[ {{\beta _1},{\beta _2},...,{\beta _G}} \right]^T}$ and ${\mathbf{A}}\left( {{{{\Psi }}_G}} \right) = \left[ {{\mathbf{a}}\left( {{\psi _1}} \right),...,{\mathbf{a}}\left( {{\psi _G}} \right)} \right] \in \mathbb{C} ^{N \times G}$. In contrast to $\bm{\beta}$, $\bm{\beta}_G \in \mathbb{C}^{G \times 1}$ denotes a grid domain coefficient vector modeled as approximately sparse, with $K$ dominant entries that provide estimates of the path directions and associated channel gains.

Using the grid approximation in (\ref{deqn_ex10}), the received observation can be approximately modeled as
\begin{equation}
	\label{deqn_ex11}
	{\mathbf{\tilde y}}  \approx {\mathbf{V}}_{}^H{\mathbf{A}}\left( {{{{\Psi }}_G}} \right){{\bm{\beta }}_G} + {\mathbf{\tilde z}}.
\end{equation}
The channel estimation problem is therefore formulated as estimating the approximately sparse vector $\bm{\beta}_G$.
Since the number of propagation paths $K$ is assumed to be known, the grid-domain coefficient vector is estimated by solving the following $\ell_0$-norm constrained least-squares problem \cite{FoucartRauhut2013}
\begin{align}
	\label{deqn_ex12}
	\mathop{\min}\limits_{{\bm{\beta}}_G}
	&\quad \left\|\mathbf{\tilde y}-\mathbf{V}^H\mathbf{A}(\Psi_G){\bm{\beta}}_G\right\|_2^2 \nonumber\\
	\mathrm{s.t.}
	&\quad \left\|{\bm{\beta}}_G\right\|_0\leq K.
\end{align}
Here, the residual accounts for both the equivalent receiver noise and the modeling error caused by the finite angular representation. The constraint limits the recovered representation to at most $K$ dominant components.
We define ${\bm{{\Gamma}}}  =  {\mathbf{V}}^H{\mathbf{A}}\left( {{{{\Psi }}_G}} \right) \in\mathbb{C}^{{N_R} \times G} $ as the measurement matrix. 

According to the basic theory of CS \cite{47}, the design of a dedicated measurement matrix based on the mutual incoherence property (MIP) contributes to improving recovery accuracy. The conditions of MIP mainly focus on the mutual coherence of the measurement matrix. The coherence $\mu$ of the measurement matrix is defined as 
\begin{equation}
	\label{deqn_ex13}
	\mu \left( {\bm{{\Gamma}}}  \right) = \mathop {\max }\limits_{i \ne j} \frac{{\left| {{\bm{{\Gamma}}} _i^H{\bm{{\Gamma}}} _j^{}} \right|}}{{{{\left\| {{\bm{{\Gamma}}} _i} \right\|}_2}{{\left\| {{\bm{{\Gamma}}} _j} \right\|}_2}}},
\end{equation}
where ${\bm{{\Gamma}}} _j^{}$ is the $j$-th column of the matrix ${\bm{{\Gamma}}} $ and ${\bm{{\Gamma}}} _i^{}$ is the $i$-th column.
A sufficient condition of a sparse reconstruction is given in \cite{51}, which is
\begin{equation}
	\label{deqn_ex14}
	\mu \left( {\bm{{\Gamma}}}  \right) < \frac{1}{{2K - 1}}. 
\end{equation} 
The Welch Bound \cite{48} provides the lower bound for $	\mu \left( {\bm{{\Gamma}}}  \right)$ as
\begin{equation}
	\label{deqn_ex15}
	\mu \left( {\bm{{\Gamma }}}  \right) \ge \sqrt {\frac{{G - {N_R}}}{{{N_R}\left( {G - 1} \right)}}},
\end{equation}
for ${\bm{{\Gamma}}}$ of size ${N_R} \times G$ and $G>N_R$.
For a fixed ${N_R}$, the Welch Bound increases with $G$. Thus, a larger $G$ improves the direction cosine resolution and reduces the grid quantization error, while it also raises the coherence lower bound, potentially degrading recovery performance \cite{49}.

In Section IV, we find that the lower bound of $\mu \left( {\bm{{\Gamma}}}  \right)$ can be reduced by a specific design of the RIMSA  response, which results in a better channel estimation performance.

\section{The Proposed TSR Method}
In this section, we propose a two-stage channel estimation method, which is based on CS theory and can enhance the accuracy of channel estimation by decreasing the lower bound on coherence of the measurement matrix. Specifically, we first adopt identical phase-shifting vectors across all RIMSAs, to reduce the column dimension of the measurement matrix via the specific antenna array manifold.  This process introduces DoA ambiguity and yields a candidate set of angle hypotheses. In the second stage, a sparse recovery procedure is used to select and refine the candidate angles associated with the physical propagation paths. Additionally, we optimize the RIMSA phase-shifting matrix for both stages.

\subsection{Reducing the Coherence of the Measurement Matrix}
It is noted that matrix  ${\mathbf{A}}\left( {{{{\Psi }}_G}} \right)$ in ($\ref{deqn_ex10}$) can be reshaped as
\begin{equation}
	\label{deqn_ex16}
	{\mathbf{A}}\left( {{{{\Psi }}_G}} \right) = \left[ {\begin{array}{*{20}{c}}
			{{{\mathbf{A}}_1}\left( {{{{\Psi }}_G}} \right)}\\
			{{{\mathbf{A}}_2}\left( {{{{\Psi }}_G}} \right)}\\
			\vdots \\
			{{{\mathbf{A}}_{{N_R}}}\left( {{{{\Psi }}_G}} \right)}
	\end{array}} \right],
\end{equation}
where ${\mathbf{A}}_{{n_r}}\left( {{{{\Psi }}_G}} \right), n_r = 1, \ldots, N_R$ is a ${{N_E} \times G}$ sub-matrix of ${\mathbf{A}}\left( {{{{\Psi }}_G}} \right)$, which
denotes the array manifold of the ${n_r}$-th RIMSA. Given that the distance between adjacent RIMSAs is ${N_E}d$, where each RIMSA comprises $N_E$ elements with inter-element spacing $d$, the phase difference between adjacent RIMSA at angle grid ${\psi _g}, g = 1,2,...,G$ is defined as 
\begin{equation}
	\label{deqn_ex17}
	\omega \left( {{\psi _g}} \right) = 2\pi \frac{{{N_E}d}}{\lambda }\sin \left( {{\psi _g}} \right).
\end{equation}
For the uniform linear array RIMSA structure as shown in Fig. \ref{fig_4}, it is not hard to see that the subarray manifold of each RIMSA satisfies
\begin{align}
	\label{deqn_ex18}
		{{\mathbf{A}}_{{n_r}}}\left( {{{{\Psi }}_G}} \right) 
		&  = {{\mathbf{A}}_1}\left( {{{{\Psi }}_G}} \right){{\mathbf{W}}_{{n_r}}},
\end{align}
where  ${{\mathbf{W}}_{{n_r}}} = \mathrm {diag}\left( {{e^{ - j\left( {{n_r} - 1} \right)\omega \left( {{\psi _1}} \right)}},...,{e^{ - j\left( {{n_r} - 1} \right)\omega \left( {{\psi _G}} \right)}}} \right)$, ${{n_r} = 1,2,...,{N_R}}$, meaning ${\mathbf{W}}_1 = {\mathbf{I}_G}$.

In the first stage, all RIMSAs are configured with the same phase-shifting vector, denoted by ${\mathbf v}_r$, i.e., ${\mathbf v}_r={\mathbf v}_1=\cdots={\mathbf v}_{N_R}$. Accordingly, the first-stage phase-shifting matrix is defined as ${\mathbf V}_r=\mathrm{blkdiag}({\mathbf v}_r,{\mathbf v}_r,\ldots,{\mathbf v}_r)$.
With this configuration, the measurement model in (\ref{deqn_ex11}) simplifies to
\begin{equation}
	\label{deqn_ex19}
	{\mathbf{\tilde y}} \approx {\mathbf{V}}_{r}^H{\mathbf{A}}\left( {{{{\Psi }}_G}} \right){{\bm{\beta }}_G} + {\mathbf{\tilde z}}.
\end{equation}
Substituting (\ref{deqn_ex16}) and (\ref{deqn_ex18}) into  ${\mathbf{A}}\left( {{{{\Psi }}_G}} \right)$ in (\ref{deqn_ex19}) yields
\begin{align}
			\label{deqn_ex20}
			{\mathbf{V}}_r^H{\mathbf{A}}\left( {{{{\Psi }}_G}} \right) &= \left[ {\begin{array}{*{20}{c}}
					{{\mathbf{v}}_r^H}&{}&{}&{}\\
					{}&{{\mathbf{v}}_r^H}&{}&{}\\
					{}&{}& \ddots &{}\\
					{}&{}&{}&{{\mathbf{v}}_r^H}
			\end{array}} \right]\left[ {\begin{array}{*{20}{c}}
					{{{\mathbf{A}}_1}\left( {{{{\Psi }}_G}} \right)}\\
					{{{\mathbf{A}}_2}\left( {{{{\Psi }}_G}} \right)}\\
					\vdots \\
					{{{\mathbf{A}}_{{N_R}}}\left( {{{{\Psi }}_G}} \right)}
			\end{array}} \right] \nonumber\\
			&= \left[ {\begin{array}{*{20}{c}}
					{{\mathbf{v}}_r^H{{\mathbf{A}}_1}\left( {{{{\Psi }}_G}} \right){{\mathbf{W}}_1}}\\
					{{\mathbf{v}}_r^H{{\mathbf{A}}_1}\left( {{{{\Psi }}_G}} \right){{\mathbf{W}}_2}}\\
					\vdots \\
					{{\mathbf{v}}_r^H{{\mathbf{A}}_1}\left( {{{{\Psi }}_G}} \right){{\mathbf{W}}_{{N_R}}}}
			\end{array}} \right]\nonumber\\
			&= {{{\mathbf{\tilde A}}}}\left( {{{{\Psi }}_G}} \right){\mathbf{U}}\left( {{{{\Psi }}_G}} \right), 
\end{align}
where 
\begin{align}
	\label{deqn_ex21}
		{{{\mathbf{\tilde A}}}}\left( {{{{\Psi }}_G}} \right) &= \left[ {\begin{array}{*{20}{c}}
				1& \cdots &1\\
				{{e^{ - j\omega \left( {{\psi _1}} \right)}}}& \cdots &{{e^{ - j\omega \left( {{\psi _G}} \right)}}}\\
				\vdots & \ddots & \vdots \\
				{{e^{ - j\left( {{N_R} - 1} \right)\omega \left( {{\psi _1}} \right)}}}& \cdots &{{e^{ - j\left( {{N_R} - 1} \right)\omega \left( {{\psi _G}} \right)}}}
		\end{array}} \right]\nonumber\\
		&= \left[ {{\mathbf{\tilde a}}\left( {\omega \left( {{\psi _1}} \right)} \right),...,{\mathbf{\tilde a}}\left( {\omega \left( {{\psi _G}} \right)} \right)} \right], 	
	\end{align}
	with ${\mathbf{\tilde a}}\left( {\omega \left( {{\psi _g}} \right)} \right) = {\left[ {1,{e^{ - j\omega \left( {{\psi _g}} \right)}},...,{e^{ - j\left( {{N_R} - 1} \right)\omega \left( {{\psi _g}} \right)}}} \right]^T}\in\mathbb{C}^{{N_R\times 1} }$,	and  ${\mathbf{U}}\left( {{{{\Psi }}_G}} \right)$ is
	\begin{equation}
		\label{deqn_ex22}
		\begin{aligned}
			{\mathbf{U}}\left( {{{{\Psi }}_G}} \right) = \mathrm {diag}\left( {u\left( {{\psi _1}} \right),...,u\left( {{\psi _G}} \right)} \right),
		\end{aligned}
	\end{equation}
with $	u\left( {{\psi _g}} \right) = {\mathbf{v}}_r^H{{\mathbf{a}}_1}\left( {{\psi _g}} \right)$. The matrix $	{{{\mathbf{\tilde A}}}}\left( {{{{\Psi }}_G}} \right)\in\mathbb{C}^{{N_R}\times{G}}$ can be viewed as the virtual array manifold of the RIMSA array with this specific phase-shifting vector. Furthermore, ${\mathbf{U}}\left( {{{{\Psi }}_G}} \right)$ represents the response of RIMSAs.
	
	Therefore, (\ref{deqn_ex19}) can be rewritten as  
	\begin{equation}
		\label{deqn_ex22a}
		{\mathbf{\tilde y}} \approx {{{\mathbf{\tilde A}}}}\left( {{{{\Psi }}_G}} \right){\mathbf{U}}\left( {{{{\Psi }}_G}} \right) {{\bm{\beta }}_G} + {\mathbf{\tilde z}}.	 
	\end{equation}	
Note that the inter-antenna phase $\omega(\psi_g)$ lies in
	\begin{equation}
		\omega \left( {{\psi _g}} \right)\in \left( { - 2\frac{{{N_E}d}}{\lambda }\pi ,2\frac{{{N_E}d}}{\lambda }\pi } \right],
	\end{equation}
	whose interval length is $2\pi Q$, where $Q= 2N_Ed/\lambda$. For $d=\lambda/2$, $Q=N_E$.
	Thus, the periodic mapping introduces a $Q$-fold DoA estimation ambiguity, and then the direction cosine grid ${{{{\Psi }}_G}}$ is partitioned into $\tilde G=G/Q$ aliasing groups. The grid points in the $\tilde g$-th, $\tilde g=1,\ldots,\tilde G$, aliasing group satisfy
	\begin{equation}
		\label{deqn_ex23}
		\zeta_{\tilde g+m\tilde G}=\zeta_{\tilde g}+\frac{2m}{Q}, \qquad m=0,\ldots,Q-1.
	\end{equation}
    Recalling that $\zeta_g=\sin(\psi_g)$, \eqref{deqn_ex23} implies that the inter-RIMSA phases associated with the corresponding physical
    angles satisfy
	\begin{equation}
		\label{deqn_add1}
		\omega(\psi_{\tilde g+m\tilde G})=\omega(\psi_{\tilde g})+2\pi m.
	\end{equation}
    Since $\mathbf{\tilde a}(\omega)$ is $2\pi$-periodic, the corresponding $Q$ columns of
    $\mathbf{\tilde A}(\Psi_G)$ are identical. The angles in the $\tilde g$-th group are defined as $\Psi_{\tilde g}=\left\{\psi_{\tilde g,q}\mid q=1,\ldots,Q\right\}$ for brevity, where $\psi_{\tilde g,q}=\psi_{\tilde g+(q-1)\tilde G}$.
	
We define $\tilde \Theta = \left\{\vartheta _1,\vartheta _2,\ldots,\vartheta _{\tilde G}\right\}$ as the set of principal phase differences between adjacent RIMSAs for these groups. Here, $\vartheta _{\tilde g}\in(-\pi,\pi]$ represents the wrapped phase difference corresponding to the $\tilde g$-th group, satisfying the congruence relation
	\begin{equation}
		\omega \left( {{\psi _{\tilde g,1}}} \right) \equiv \cdots \equiv \omega \left( {{\psi _{\tilde g,Q}}} \right) \equiv {\vartheta _{\tilde g}} \pmod{2\pi}.
	\end{equation}
The mapping relationship between the principal phase ${{{\tilde \Theta }}}$ and the physical angles ${{{{\Psi }}_G}}$ is given by
	\begin{equation}
		\label{deqn_ex24}
		\Psi _{\tilde g} = \left\lbrace  \arcsin \left( {\frac{{\lambda \left( {{\vartheta _{\tilde g}} + 2\pi i} \right)}}{{2\pi {N_E}d}}} \right) \middle| -\frac{Q}{2} - \frac{\vartheta _{\tilde g}}{2\pi} < i \le \frac{Q}{2} - \frac{\vartheta _{\tilde g}}{2\pi} \right\rbrace.
	\end{equation}
Accordingly, after merging the $Q$ identical columns associated with each $\Psi_{\tilde g}$ in ${{{\mathbf{\tilde A}}}}\left( {{{{\Psi }}_G}} \right)$, Eq.~(\ref{deqn_ex19}) can be expressed as
	\begin{equation}
		\label{deqn_ex25}
		\begin{aligned}
			{\mathbf{\tilde y}} \approx {\mathbf{\tilde A}}\left( {{{\tilde \Theta }}} \right){{\bm{\beta }}_{\tilde G}} + {\mathbf{\tilde z}},
		\end{aligned} 
	\end{equation}
	where ${\mathbf{\tilde A}}\left( {{{\tilde \Theta }}} \right) = \left[ {{\mathbf{\tilde a}}\left( {{\vartheta _1}} \right),...,{\mathbf{\tilde a}}\left( {{\vartheta _{\tilde G}}} \right)} \right] \in\mathbb{C}^{{N_R} \times \tilde G} $ with ${\mathbf{\tilde a}}\left( {{\vartheta _{\tilde g}}} \right) = {\left[ {1,{e^{ - j{\vartheta _{\tilde g}}}},...,{e^{ - j\left( {{N_R} - 1} \right){\vartheta _{\tilde g}}}}} \right]^T}$. 
	
After merging the identical columns in ${{{\mathbf{\tilde A}}}}\left( {{{{\Psi }}_G}} \right)$, we define ${{\bm{\beta }}_{\tilde G}}$ as the complex attenuation according to ${{{{\Psi }}_G}} $, which is ${{\bm{\beta }}_{\tilde G}}   =  {\left[ {{\beta _{\tilde 1}},{\beta _{\tilde 2}},...,{\beta _{\tilde G}}} \right]^T}$ and can be calculated by 
	\begin{equation}
		\label{deqn_ex26}
		{{{\beta }}_{\tilde g}} = \left[ {u\left( {{\psi _{\tilde g,1}}} \right),u\left( {{\psi _{\tilde g,2}}} \right),...,u\left( {{\psi _{\tilde g,Q}}} \right)} \right]\left[ {\begin{array}{*{20}{c}}
				{{\beta _{\tilde g,1}}}\\
				{{\beta _{\tilde g,2}}}\\
				\vdots \\
				{{\beta _{\tilde g,{Q}}}}
		\end{array}} \right], 
	\end{equation}
	where $u\left( {{\psi _{\tilde g,q}}} \right) = {\mathbf{v}}_r^H{{\mathbf{a}}_1}\left( {{\psi _{\tilde g,q}}} \right)$ and ${\beta _{\tilde g,q}}$ is the channel complex attenuation, which is associated to ${\psi _{\tilde g,q}}, {\tilde g = 1,2,...,\tilde G}, q = 1,2,...,{Q}$.

According to the reduced observation model in \eqref{deqn_ex25}, the first-stage sparse recovery problem \eqref{deqn_ex12} becomes
\begin{align}
	\label{deqn_ex27}
	\mathop{\min}\limits_{{\bm{\beta}}_{\tilde G}}
	&\quad \left\|\mathbf{\tilde y}-\tilde{\bm{\Gamma}}{\bm{\beta}}_{\tilde G}\right\|_2^2 \nonumber\\
	\mathrm{s.t.}
	&\quad \left\|{\bm{\beta}}_{\tilde G}\right\|_0\leq K,
\end{align}
where the measurement matrix now becomes $\tilde {\bm{{\Gamma}}} ={\mathbf{\tilde A}}\left( {{{\tilde \Theta }}} \right){ \in\mathbb{C} ^{{N_R} \times \tilde G}}$. The DGMP algorithm in \cite{41} is employed as a greedy approximate solver for this $\ell_0$-norm constrained problem. It first identifies a coarse atom and then locally refines the corresponding principal phase. Consequently, the refined phase estimates are not restricted to the predefined grid $\tilde\Theta$, which effectively mitigates the basis mismatch error caused by finite angular discretization.
Compared to that of ${\bm{{\Gamma}}} $ in (\ref{deqn_ex15}), the lower bound of the coherence of the measurement matrix  $\tilde {\bm{{\Gamma}}} $ now becomes 
	\begin{equation}
		\label{deqn_ex28}
		\mu \left( \tilde {\bm{{\Gamma}}}  \right) \ge \sqrt {\frac{{{\tilde G} - {N_R}}}{{{N_R}\left( {{\tilde G} - 1} \right)}}},
	\end{equation}
which is smaller than the lower bound of $\mu \left( {\bm{{\Gamma}}}  \right)$ in ($\ref{deqn_ex15}$). This will improve the performance of sparse signal reconstruction under the same on-grid density. 

In short, we exploit the property that RIMSA array manifolds can be merged by the special design using identical phase-shifting vectors across all RIMSAs in this stage to decrease the coherence of the measurement matrix, thereby achieving high accuracy in solving ($\ref{deqn_ex27}$). Then, we can get $K$ dominant components together with the refined principal phase estimates $\left\{ {{{\hat \vartheta }_k},k = 1,2,...,K} \right\}$. Notably, if multiple physical paths collapse into the same aliasing group, the DGMP algorithm still extracts $K$ components, thereby inevitably introducing spurious phase estimates into the output set. Moreover, due to the local grid refinement in DGMP, $\hat \vartheta_k$ is not restricted to the original coarse grid ${\tilde \Theta }$.
	
However, ambiguous angles are also introduced in this stage. Specifically, for each refined principal phase ${{\hat \vartheta }_k}$, $k=1,2,\ldots,K$, (\ref{deqn_ex24}) yields $Q$ corresponding physical angles ${{\psi}_{k,q}}$ in $(-\pi/2,\pi/2]$. When ${{\hat \vartheta }_k}$ lies on the coarse grid, these angles coincide with the corresponding points in ${{\Psi }}_G$. After local DGMP refinement, they are generally off-grid. They form the candidate set ${{\bar \Psi }_{\bar G}} = \left\{ {{\psi _{k,q}},k = 1,\ldots,K,q = 1,\ldots,Q} \right\}$, 
which is expected to contain candidate angles sufficiently close to the corresponding physical DoAs.
Even if path collisions cause some spurious $\hat{\vartheta}_k$ and false candidate angles, the $Q$ candidate angles derived from the correct $\hat{\vartheta}_k$ still include all physical paths folded into this group. Therefore, the candidate set $\bar{\Psi}_{\bar{G}}$ contains all expected paths, which enables the sparse recovery in the second stage.
Therefore, our next stage is to refine the DoAs from the $\bar G = {Q}K$ angles of the set ${{\bar \Psi }_{\bar G}}$. Before proceeding, we first detail the construction of ${{\mathbf{V}}_r}$ in this stage.

	\subsection{The Design of ${{\mathbf{V}}_r}$}
	In this section, the detailed design of ${{\mathbf{V}}_r}$ is considered. 
	Under the assumption that ${\mathbf{v}}_r^{}={\mathbf{v}}_1^{}  =  \cdots  = {\mathbf{v}}_{{N_R}}$, according to ($\ref{deqn_ex19}$) and ($\ref{deqn_ex20}$), the power contributed by the $k$-th physical path arriving from ${\theta _k}$, $k=1,\ldots,K$, can be written as
		\begin{align}	\label{deqn_ex29}
				P\left( {{\theta _k},{\mathbf{v}}_r} \right)& ={\mathbf{\tilde a}}^{H}\left( {\omega \left( {{\theta _k}} \right)} \right){{\beta }}_k^{H} u^H\left( {{\theta _k}} \right)  u\left( {{\theta _k}} \right) {{\beta }}_k^{}  {\mathbf{\tilde a}}\left( {\omega \left( {{\theta _k}} \right)} \right) \nonumber
				\\
				&= {\mathbf{\tilde a}}^{H}\left( {\omega \left( {{\theta _k}} \right)} \right){{\beta }}_k^H\tilde P\left( {{\theta _k},{\mathbf{v}}_r^{}} \right){{\beta }}_k^{}{\mathbf{\tilde a}}\left( {\omega \left( {{\theta _k}} \right)} \right),
			\end{align}
		where ${{\beta }}_k$ and ${\mathbf{\tilde a}}^{H}\left( {\omega \left( {{\theta _k}} \right)} \right)$ are fixed for a given physical path angle ${\theta _k}$, and $\tilde P\left( {{\theta _k},{\mathbf{v}}_r^{}} \right) = u^H\left( {{\theta _k}} \right) u\left( {{\theta _k}} \right)$.
		
\begin{figure}
				\centering
				\subfloat[]{
					\includegraphics[width=0.8\linewidth]{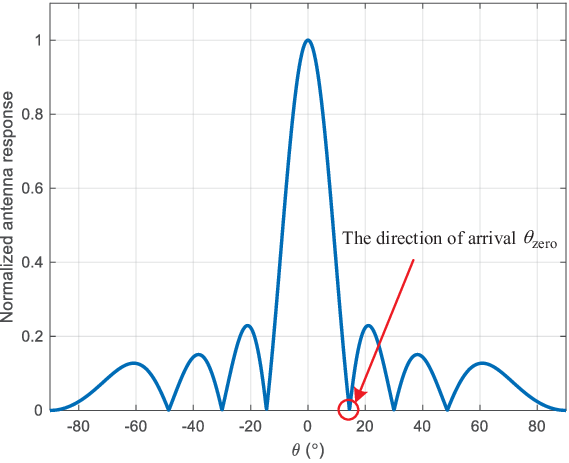} 
				}\\
				\subfloat[]{
					\includegraphics[width=0.8\linewidth]{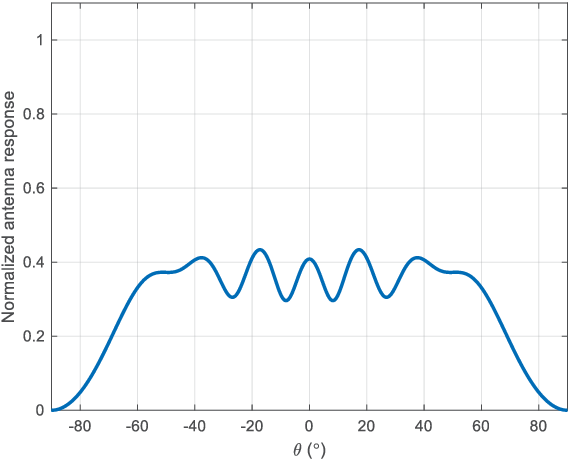}
				}
				\caption{(a) Normalized metasurface antenna response for RIMSA with ${\mathbf{v}}_r={\mathbf{1}}$; (b) Normalized metasurface antenna response for RIMSA with optimized ${\mathbf{v}}_r$ via (\ref{deqn_ex30}).}
				\label{fig_5}
			\end{figure}

		According to (\ref{deqn_ex22}), $u\left( {{\theta _k}} \right)={\mathbf{v}}_r^H{{\mathbf{a}}_1}\left( {{\theta _k}} \right)$ is determined by ${\mathbf{v}}_r$ for a given ${\theta _k}$.
		If ${\mathbf{v}}_r^H{{\mathbf{a}}_1}\left( {{\theta _k}} \right)$ is small, the received power of the $k$-th physical path is greatly attenuated, resulting in seriously deteriorated channel estimation performance. For example, Fig. \ref{fig_5}(a) depicts the normalized RIMSA response in $\left( { - \pi /2,\pi /2} \right]$ with ${N_E} =8$ and ${\mathbf{v}}_r = \mathbf{1}_{N_E}\in\mathbb{C}^{N_E\times 1}$. The normalized radiation pattern exhibits multiple nulls as shown in Fig. \ref{fig_5}(a), and we use ${\theta _{zero,i}}$ to represent the $i$-th null. Signals arriving from ${\theta _{zero,i}}$ experience near-zero received power, making angle recovery infeasible.

		In order to deal with this issue, we should design the phase-shifting vector ${\mathbf{v}}_r$ to satisfy some constraint in the whole interested DoA range where the impinging signals may come from, i.e., cell sector coverage. Because the physical DoAs $\{\theta_k\}_{k=1}^{K}$ are unknown during this offline design, we discretize the interested DoA range using a dense design set $\Phi_M=\{\phi_m\mid m=1,\ldots,M\}$, which is independent of the CS grid $\Psi_G$. This problem could be modeled as an optimization problem to restrain the antenna response to avoid the nulls in the covered DoA range, which could be expressed as
			\begin{align}
				\label{deqn_ex30}
				{\mathop {\min }\limits_{{\gamma },{\mathbf{v}}_r^{}} }&~~{\sum\limits_{m = 1}^M {{{\left| {{\mathbf{a}}_1^H\left( {{\phi _m}} \right){\mathbf{v}}_r^{}{\mathbf{v}}_r^H{{\mathbf{a}}_1}\left( {{\phi _m}} \right) -{\gamma } {P_m}} \right|}^2}} }  \nonumber \\ 
				\mathrm {s.t.}
				&~~~~~~~{\left| {{v_{{r},{n_e}}}} \right| = 1,{n_e} = 1,...,{N_E}},
			\end{align}
			where ${P_m}$ is the desired antenna response power at ${{\phi _m}}$. Since $\Phi_M$ samples only the interested DoA range, ${P_m}=1$ for all $m=1,\ldots,M$, and $\gamma$ is a real scaling factor.

			The optimization variables in (\ref{deqn_ex30}) are scaling factor $\gamma $ and the phase-shifting vector ${{\mathbf{v}}_r}$. 
			For a given $\mathbf{v}_r$, the optimal  $\gamma$ can be derived  in closed form. Let $h(\gamma,\mathbf{v}_r)$ denote the objective function in (\ref{deqn_ex30}). Since it is quadratic in the real-valued variable $\gamma$, its derivative is
			\begin{align}
				\label{deqn_ex32}
					\frac{{\partial h\left( {\gamma ,{\mathbf{v}}_r} \right)}}{{\partial \gamma }} &=  - 2\sum\limits_{m = 1}^M {\left( {{\mathbf{a}}_1^H\left( {{\phi _m}} \right){\mathbf{v}}_r^{}{\mathbf{v}}_r^H{{\mathbf{a}}_1}\left( {{\phi _m}} \right)} \right)}P_m  \nonumber \\
					 &\quad
					 + 2\gamma \sum\limits_{m = 1}^M {P_m^2}, 
			\end{align}
		which yields the optimal $\gamma$ as
		\begin{equation}
			\label{deqn_ex33}
			\gamma^* = \sum\limits_{m = 1}^M \left( {{\mathbf{a}}_1^H\left( {{\phi _m}} \right){\mathbf{v}}_r^{}{\mathbf{v}}_r^H{{\mathbf{a}}_1}\left( {{\phi _m}} \right)} \right) P_m \bigg/ \sum\limits_{m = 1}^M {P_m^2}.
		\end{equation}
		With the optimal $\gamma^*$, which is also a quadratic function of ${{\mathbf{v}}_r^{}}$, the next step is to find the optimal ${{\mathbf{v}}_r^{}}$. Substituting $\gamma^*$ in (\ref{deqn_ex33}) to (\ref{deqn_ex30}) transforms the problem into minimizing a fourth-order polynomial of ${{\mathbf{v}}_r^{}}$. The constraint for ${{\mathbf{v}}_r^{}}$ is non-convex, which is a non-trivial problem. 
				
		Note that $v_{{r},{n_e}}^{} =  {e^{ - j\alpha _{{r},{n_e}}^{}}}$ and ${{\mathbf{v}}_r^{}}$ can be written as ${\mathbf{v}}_r^{}\left( {\bm{\alpha }} \right)$ according to (\ref{deqn_ex1}), where ${\bm{\alpha }} = {\left[ {{\alpha _{r,1}},...,{\alpha _{r,{N_E}}}} \right]^T}$, so we can take ${\bm{\alpha }}$ as the optimization variable. 
				Therefore, the optimization problem (\ref{deqn_ex30}) can be written as
				\begin{equation}
					{\mathop {\min }\limits_{{\bm{\alpha }}} }\ {\sum\limits_{m = 1}^M {{{\left| {{\mathbf{a}}_1^H\left( {{\phi _m}} \right){\mathbf{v}}_r^{}\left( {\bm{\alpha }} \right){\mathbf{v}}_r^H\left( {\bm{\alpha }} \right){{\mathbf{a}}_1}\left( {{\phi _m}} \right) - {\gamma^*} {P_m}} \right|}^2}} }.	\label{deqn_ex31_a}
				\end{equation}
				To simplify the objective function, we rewrite the scalar terms as quadratic forms with respect to ${\mathbf{v}}_r^{}\left( {\bm{\alpha }} \right)$. Specifically, the first term becomes ${\mathbf{v}}_r^H\left( {\bm{\alpha }} \right){\mathbf{a}}_1^{}\left( {{\phi _m}} \right){\mathbf{a}}_1^H\left( {{\phi _m}} \right){\mathbf{v}}_r^{}\left( {\bm{\alpha }} \right)$. Then, substituting $\gamma^*$ from (\ref{deqn_ex33}) into the second term yields
					\begin{align}
						{\gamma ^*}{P_m} &= {P_m}\frac{{\sum\limits_{\ell = 1}^M {{P_\ell}\left( {{\mathbf{v}}_r^H\left( {\bm{\alpha }} \right){{\mathbf{a}}_1}\left( {{\phi _\ell}} \right){\mathbf{a}}_1^H\left( {{\phi _\ell}} \right){\mathbf{v}}_r^{}\left( {\bm{\alpha }} \right)} \right)} }}{{\sum\limits_{\ell = 1}^M {P_\ell^2} }} \nonumber\\
						&= {\mathbf{v}}_r^H\left( {\bm{\alpha }} \right)\left[ {{P_m}\frac{{\sum\limits_{\ell = 1}^M {{P_\ell}{{\mathbf{a}}_1}\left( {{\phi _\ell}} \right){\mathbf{a}}_1^H\left( {{\phi _\ell}} \right)} }}{{\sum\limits_{\ell = 1}^M {P_\ell^2} }}} \right]{\mathbf{v}}_r^{}\left( {\bm{\alpha }} \right).
					\end{align}
					By extracting the common vectors ${\mathbf{v}}_r^H\left( {\bm{\alpha }} \right)$ and ${\mathbf{v}}_r^{}\left( {\bm{\alpha }} \right)$, the problem can be further transformed to
				\begin{equation}
					{\mathop {\min }\limits_{\bm{\alpha }} }\ {\sum\limits_{m = 1}^M {{{\left| {{\mathbf{v}}_r^H\left( {\bm{\alpha }} \right){{\mathbf{B}}_m}{\mathbf{v}}_r^{}\left( {\bm{\alpha }} \right)} \right|}^2}} },	\label{deqn_ex31_b}
				\end{equation}
				where $	{{\mathbf{B}}_m}= {\mathbf{a}}_1^{}\left( {{\phi _m}} \right){\mathbf{a}}_1^H\left( {{\phi _m}} \right) - {P_m}\frac{\sum\limits_{\ell = 1}^M{P_\ell} {\left( {{{\mathbf{a}}_1}\left( {{\phi _\ell}} \right){\mathbf{a}}_1^H\left( {{\phi _\ell}} \right)} \right)}}{{\sum\limits_{\ell = 1}^M {P_\ell^2} }} $.

				The optimization problem (\ref{deqn_ex31_b}) now becomes an unconstrained fourth
				order trigonometric polynomial minimization problem. This minimization problem can be solved by the Limited-Memory Broyden–Fletcher–Goldfarb and Shanno (L-BFGS) algorithm \cite{60}, which is a popular quasi-Newton algorithm and was described in detail in \cite{57}. Unlike Newton-type approaches, which require positive definite Hessians and exact second derivatives, L-BFGS approximates curvature information through gradient updates, ensuring both computational tractability and robustness to non-convexity.  
				
				To implement the L-BFGS algorithm, we need to calculate the gradient of the objective function and obtain the changes in gradients, which contributes to construction of a superlinear convergence model. For this purpose, we first transform  the objective function of (\ref{deqn_ex31_b})  into a matrix form as
				\begin{equation}
					\label{deqn_ex31_d}
					\begin{array}{*{20}{c}}
						{\mathop {\min }\limits_{\bm{\alpha }} }\mathcal{J}(\bm{\alpha})
					\end{array},
				\end{equation}
				where $\mathcal{J}(\bm{\alpha})={{{\mathbf{x}}^H}\left( {\bm{\alpha }} \right){\mathbf{Bx}}\left( {\bm{\alpha }} \right)}$ with $	{\mathbf{x}}\left( {\bm{\alpha }} \right) = {\mathrm{vec}}\left( {{\mathbf{v}}_r^{}\left( {\bm{\alpha }} \right){\mathbf{v}}_r^H\left( {\bm{\alpha }} \right)} \right)$, $	{{\mathbf{b}}_m} = {\mathrm{vec}}\left( {{{\mathbf{B}}_m}} \right)$, and $	{\mathbf{B}} = \sum\limits_{m = 1}^M {{{\mathbf{b}}_m}{\mathbf{b}}_m^H}$. Here, ${\mathrm{vec}}\left(  \cdot  \right)$ vectorizes  a matrix by stacking its columns. Then, we can compute the gradient of $\mathcal{J}(\alpha)$ as follows:
				\begin{equation}
					\label{deqn_ex31_e}
					\frac{{\partial \mathcal{J}(\alpha)}}{{\partial {\bm{\alpha }}}} = 2{\mathop{\mathrm {Re}}\nolimits} \left( {{{\left[ {\frac{{\partial {\mathbf{x}}\left( {\bm{\alpha }} \right)}}{{\partial {\bm{\alpha }}}}} \right]}^H}{\mathbf{Bx}}\left( {\bm{\alpha }} \right)} \right),
				\end{equation}
				where
				\begin{equation}
					\label{deqn_ex31_f}
					\frac{{\partial {\mathbf{x}}\left( {\bm{\alpha }} \right)}}{{\partial {\alpha _{r,{n_e}}}}} = {\mathrm{vec}}\left( {\frac{{\partial {{\mathbf{v}}_r}\left( {\bm{\alpha }} \right)}}{{\partial {\alpha _{r,{n_e}}}}}{\mathbf{v}}_r^H\left( {\bm{\alpha }} \right) + {\mathbf{v}}_r^{}\left( {\bm{\alpha }} \right)\frac{{\partial {\mathbf{v}}_r^H\left( {\bm{\alpha }} \right)}}{{\partial {\alpha _{r,{n_e}}}}}} \right).
				\end{equation}
				The detailed L-BFGS algorithm is shown in Algorithm \ref{alg:BFGS}.
				
				\begin{algorithm}[t]
					\caption{L-BFGS Algorithm for optimizing ${\mathbf{v}}_r$  }\label{alg:BFGS}
					\begin{algorithmic}[1]
						\REQUIRE: Initialize the number of L-BFGS updates, $\eta $. Choose initial value ${\bm{\alpha }}_1$ and calculate $\mathcal{J}\left( {\bm{\alpha }_1} \right)$ and $\frac{{\partial \mathcal{J}\left( {\bm{\alpha }}_1 \right)}}{{\partial {\bm{\alpha }}_1}}$ using Eq. (\ref{deqn_ex31_d}) and Eq. (\ref{deqn_ex31_e}), respectively.
						\STATE Set $p=1$, $\nabla {\mathcal{J}_1} = \frac{{\partial \mathcal{J}\left( {{{\bm{\alpha }}_1}} \right)}}{{\partial {{\bm{\alpha }}_1}}}$, and ${{\mathbf{w}}_1} =  - \nabla {\mathcal{J}_1}$.
						\REPEAT
						\STATE Calculate search step size ${\rho  _p}$ according to the specific line search rule.  
						\STATE Calculate ${{\mathbf{u}}_p} = {\rho _p}{{\mathbf{w}}_p}$ and ${{\bm{\alpha }}_{p + 1}} = {{\bm{\alpha }}_p} + {\rho _p}{{\mathbf{w}}_p}$.
						\STATE Calculate $\nabla {\mathcal{J}_{p+1}}$ and ${{\mathbf{d}}_p} = \nabla {\mathcal{J}_{p+1}}-\nabla {\mathcal{J}_{p}}$.
						\STATE Set ${\mathbf{q}}=\nabla {\mathcal{J}_{p+1}}$.
						\FOR{$i=p, p-1,\dots, \max(1, p-\eta+1)$} 
						\STATE ${t_i} = \frac{{{\mathbf{u}}_i^H{\mathbf{q}}}}{{{\mathbf{d}}_i^H{{\mathbf{u}}_i}}},{\mathbf{q}} = {\mathbf{q}} - {t_i}{\mathbf{d}}_i^{}$.
						\ENDFOR 
						\STATE $\mathbf{r} = \frac{\mathbf{u}_p^H \mathbf{d}_p}{\mathbf{d}_p^H \mathbf{d}_p} \mathbf{q}$.
						\FOR{$i=\max(1, p-\eta+1),p-\eta+2,\dots,p$} 
						\STATE $\kappa  = \frac{{{\mathbf{d}}_i^H{\mathbf{r}}}}{{{\mathbf{d}}_i^H{\mathbf{u}}_i^{}}},{\mathbf{r}} = {\mathbf{r}} + \left( {{t_i} - \kappa } \right){\mathbf{u}}_i^{}$.
						\ENDFOR 
						\STATE ${{\mathbf{w}}_{p+1}} =  - {\mathbf{r}}$, $p=p+1$.
						\UNTIL {The convergence condition is satisfied.}
					\end{algorithmic}
				\end{algorithm}
				
				The parameter vector ${{\bm{\alpha}}}$ is iteratively updated until the convergence of the objective function (\ref{deqn_ex31_d}), yielding an optimized phase-shifting vector 
				${{\mathbf{v}}_r^{}}$ that guarantees robustness of RIMSA response.
				For example, consider a RIMSA system with ${N_E}=8$ metamaterial elements and the interested DoA range is $\left[ { - \frac{1}{3}\pi , \frac{1}{3}\pi } \right]$, which is a typical cell sector coverage of 120 degrees. Solving (\ref{deqn_ex30}) yields the normalized antenna response shown in Fig. \ref{fig_5}(b), where no pattern null exists within the entire interested DoA range and the RIMSA response remains nearly constant.
				
				{\emph {Remark 1}:}
				The phase-shifting vector ${{\mathbf{v}}_r}$ depends exclusively on the design angles $\{\phi_m\}_{m=1}^{M}$ sampled within the target DoA range, irrespective of any channel realization.
				Therefore, to reduce the computational complexity, we can pre-optimize ${{\mathbf{v}}_r}$ once off-line, which is independent of the received signals.

\subsection{Physical DoA Selection from the Candidate DoA Set}
According to the discussion in Section IV-A, we have obtained ${{\bar \Psi }_{\bar G}}$ containing $\bar G$ candidate DoAs, including $K$ candidate DoAs sufficiently close to the $K$ true DoAs, respectively. Consequently, the original channel model in (\ref{deqn_ex10}) can be approximately reformulated as
\begin{equation}
	\label{deqn_ex35_a}
	{\mathbf{h}} \approx {\mathbf{A}}\left( {{{{{\bar \Psi }}}_{\bar G}}} \right){{\bm{\bar \beta }}_{\bar G}} ,
\end{equation}
where ${\mathbf{A}}\left( {{{{{\bar \Psi }}}_{\bar G}}} \right)\in\mathbb{C}^{{N}\times{\bar G}}$, and ${{\bm{\bar \beta }}_{\bar G}}\in\mathbb{C}^{\bar G \times 1}$ is modeled as a sparse coefficient vector with at most $K$ dominant nonzero entries. 
This ${{\bm{\bar \beta }}_{\bar G}}$ vector can be specifically represented as
\begin{equation}
	\label{deqn_ex36}
	{{{\bm{\bar \beta }}}_{\bar G}} = {\left[ {{{{\bm{\bar \beta }}}_1}^{T},...,{{{\bm{\bar \beta }}}_K^{T}}} \right]^T},
\end{equation}
with ${{{\bm{\bar \beta }}}_k} = \left[ {{{ \beta }_{k,1}},...,{{ \beta }_{k,Q}}} \right]^{T}\in\mathbb{C}^{Q\times 1}$ and ${\beta _{k,q}}$ is the complex channel attenuation, which is corresponding to ${\psi _{k,q}, k=1,\ldots, K, q=1,\ldots, Q}$. Given element spacing $d= \lambda /2$, we note that $Q={N_E}$ and ${\bar G}={N_E}K$. Because each RIMSA is equipped with numerous metamaterial elements, which means the resulting steering matrix ${\mathbf{A}}\left( {{{{{\bar \Psi }}}_{\bar G}}} \right)$ is generally full column rank.

In the second stage, we consider how to select $K$ optimal channel attenuation coefficients and the corresponding angles from ${{\bm{\bar \beta }}_{\bar G}}$ and ${{\bar \Psi }_{\bar G}}$. The parameter selection process is fundamentally equivalent to sparse signal recovery of ${{\bm{\bar \beta }}_{\bar G}}$. 
In contrast to the first stage where all RIMSAs employ an identical phase-shifting vector to preserve the aliasing structure, the RIMSAs in this stage are allowed to use different phase-shifting vectors, thereby breaking the aliasing ambiguity and distinguishing the candidate physical DoAs.
We denote the second-stage RIMSA phase-shifting matrix by  ${\mathbf{V}}_f = {\mathrm {blkdiag}}\left( {{\mathbf{v}}_{f,1}^{},{\mathbf{v}}_{f,2}^{},...,{\mathbf{v}}_{f,{N_R}}^{}} \right)$ and then use it to receive pilot signal again, where ${\mathbf{v}}_{f,{n_r}}^{},{n_r} = 1, \ldots ,{N_R}$ is the phase-shifting vector of ${n_r}$-th metasurface antenna in the form of ${\mathbf{v}}_{f,{n_r}}^{} = 1/\sqrt{N_E}{{\left[ {{v_{n_r,1}},...,{v_{n_r,N_E}}} \right]}^T},{n_r} = 1,2,...,{N_R}$ and $\left| {{v_{{n_r,n_e}}}} \right| = 1,{n_e} = 1,...,{N_E}$. 

Similar to (\ref{deqn_ex7}) and (\ref{deqn_ex9}), the received pilot signals can be approximately expressed as 
\begin{equation}
	\label{deqn_ex35}
	{\mathbf{\bar y}} \approx {\mathbf{V}}_f^H{\mathbf{A}}\left( {{{{{\bar \Psi }}}_{\bar G}}} \right){{\bm{\bar \beta }}_{\bar G}} + {\mathbf{\bar z}} ,
\end{equation}
and we can transform the channel estimation problem into the following optimization problem
\begin{align}
	\label{deqn_ex37}
	\mathop{\min}\limits_{{\bm{\bar\beta}}_{\bar G}}
	&\quad \left\|\mathbf{\bar y}-\bar{\bm{\Gamma}}{\bm{\bar\beta}}_{\bar G}\right\|_2^2 \nonumber\\
	\mathrm{s.t.}
	&\quad \left\|{\bm{\bar\beta}}_{\bar G}\right\|_0\leq K,
\end{align}
where $\bar {\bm{{\Gamma}}}  =  {\mathbf{V}}_f^H{\mathbf{A}}\left( {{{{{\bar \Psi }}}_{\bar G}}} \right){ \in\mathbb{C} ^{{N_R} \times {\bar G}}}$ is the current measurement matrix. This sparse recovery problem can be approximately solved using the OMP  algorithm.

To further improve accuracy, we can design ${\mathbf{V}}_f^H$ to minimize the current coherence $\mu\left( {\bar {\mathbf{\Gamma}  }} \right)$. Based on the analysis of \cite{50} and \cite{51}, we can design a measurement matrix with minimum coherence by the following optimization problem
\begin{align}
	\label{deqn_ex38}
	{\mathop {\min }\limits_{{\mathbf{V}}_f} } 
	&~~{\left\| {{{\mathbf{A}}^H}\left( {{{{{\bar \Psi }}}_{\bar G}}} \right){\mathbf{V}}_f^{}{\mathbf{V}}_f^H{\mathbf{A}}\left( {{{{{\bar \Psi }}}_{\bar G}}} \right) - {{\mathbf{I}}_{\bar G}}} \right\|_\mathrm{F}^2},	 \nonumber\\
	\mathrm {s.t.}
	&~~{{\mathbf{V}}_f = {\mathrm {blkdiag}}\left( {{\mathbf{v}}_{f,1}^{},{\mathbf{v}}_{f,2}^{},...,{\mathbf{v}}_{f,{N_R}}^{}} \right)}, \nonumber\\
	&~~{\left| {{v_{{n_r},{n_e}}}} \right| = 1,{n_r} = 1,...,{N_R},{n_e} = 1,...,{N_E}}.
\end{align}
The optimization problem in (\ref{deqn_ex38}) minimizes the sum of squared inner products of the columns of the measurement matrix. The phase-shifting matrix must satisfy both constant-modulus constraints and block-diagonal structural constraints, resulting in a non-convex problem. Fortunately, it can be solved using a Riemannian manifold algorithm, which has been discussed in many works to deal with a series of constant-modulus constraints related to phase shifters such as \cite{54}. This algorithm considers the optimization problem in the Riemannian manifold space, which is defined by constant-modulus restrictions.  Riemannian gradient will be calculated and  updated iteratively in this direction. Finally, in order to satisfy the constant modulus constraints, the updated optimization values are retracted into Riemannian manifold space.

	\begin{algorithm}[t]
	\caption{Riemannian manifold algorithm for optimizing ${\mathbf{V}}_f$ }\label{alg:RMO}
	\begin{algorithmic}[1]
		\REQUIRE ${\mathbf{A}}\left( {{{{{\bar \Psi }}}_{\bar G}}} \right)$, the initial point ${\mathbf{V}}_f^{\left( 0 \right)}$.
		\STATE Set $i=0$ and calculate ${{\mathbf{D}}_0}=-gra{d_{{\mathbf{V}}_f^{\left( 0 \right)}}}f$.
		\REPEAT
		\STATE Obtain Armijo backtracking line search step size ${\tau _i}$ according to \cite{54}.  
		\STATE Find the new point by ${\mathbf{V}}_f^{\left( i+1 \right)} = \exp\left( {j\angle \left( {{\mathbf{V}}_f^{\left( i \right)} + {\tau _i}{{\mathbf{D}}_i}} \right)} \right) \odot {\mathbf{C}}$.
		\STATE Calculate the vector transfer factor by ${\mathbf D}_i^{\text{trans}} = {\mathbf D}_i - {\mathrm {Re}}\left\{ {\mathbf D}_i \odot {{\mathbf V}_f^{(i+1)}}^* \right\} \odot {\mathbf V}_f^{(i+1)}$.
		\STATE Calculate $gra{d_{{\mathbf{V}}_f^{\left( i+1 \right)}}}f$ and then get Polak-Ribière parameter ${\xi _{i + 1}}$ according to \cite{54}.
		\STATE Calculate ${\mathbf{D}}_{i + 1}^{} =  - gra{d_{{\mathbf{V}}_f^{\left( i+1 \right)}}}f + {\xi _{i + 1}}{\mathbf{D}}_i^{\text{trans}}$.
		\STATE $i = i + 1$.
		\UNTIL{The convergence condition is satisfied.}
	\end{algorithmic}
\end{algorithm}

According to the theory of Riemannian manifold algorithm, we need to calculate the Riemannian gradient first. We define
\begin{equation}
	\label{deqn_ex45}
	f = {\left\| {{{\mathbf{A}}^H}\left( {{{{{\bar \Psi }}}_{\bar G}}} \right){\mathbf{V}}_f^{}{\mathbf{V}}_f^H{\mathbf{A}}\left( {{{{{\bar \Psi }}}_{\bar G}}} \right) - {{\mathbf{I}}_{\bar G}}} \right\|_\mathrm{F}^2},
\end{equation}
and Riemannian manifold space
\begin{equation}
	\label{deqn_ex46}
	{\cal V} = \left\{ {{\mathbf{V}}_f^{\left( i \right)} = {\mathrm {blkdiag}}\left( {{{\mathbf{v}}_{f,1}}^{\left( i \right)},{{\mathbf{v}}_{f,2}}^{\left( i \right)},...,{{\mathbf{v}}_{f,{N_R}}^{\left( i \right)}}} \right)} \right\},
\end{equation}
where ${\mathbf{V}}_f^{\left( i \right)}$ is the $i$-th iteration value.
Based on the concept of tangent space, Riemannian gradient can be calculated as the orthogonal projection of the traditional Euclidean gradient onto the Riemannian tangent space. We denote $gra{d_{{\mathbf{V}}_f^{\left( i \right)}}}f$ to be the Riemannian gradient of ${\mathbf{V}}_f^{\left( i \right)}$, which is 
\begin{equation}
	\label{deqn_ex48}
	gra{d_{{\mathbf{V}}_f^{\left( i \right)}}}f = {Grad _{{\mathbf{V}}_f^{\left( i \right)}}}f - {\mathop{\mathrm {Re}}\nolimits} \left\{ {{Grad _{{\mathbf{V}}_f^{\left( i \right)}}}f \odot {{\left( {{\mathbf{V}}_f^{\left( i \right)}} \right)}^*}} \right\} \odot {\left( {{\mathbf{V}}_f^{\left( i \right)}} \right)},
\end{equation}	
where $ \odot $ is Hadamard product, ${\mathop{\mathrm {Re}}\nolimits}\left\{ \cdot \right\}$ represents the real part and the Euclidean gradient ${Grad _{{\mathbf{V}}_f^{\left( i \right)}}}f$ can be expressed as
	\begin{equation}
	\label{deqn_ex50}
	Gra{d_{{\mathbf{V}}_f^{\left( i \right)}}}f = {\nabla _{{\mathbf{V}}_f^{\left( i \right)}}}f \odot {\mathbf{C}},
\end{equation}
	where $\mathbf C=\operatorname{blkdiag}(\mathbf 1_{N_E},\ldots,\mathbf 1_{N_E})\in\mathbb C^{N\times N_R}$
	is a block diagonal matrix with the same structure as ${\mathbf{V}}_f^{\left( i \right)}$, and ${\nabla _{{\mathbf{V}}_f^{\left( i \right)}}}f$ is given by
\begin{align}
	{\nabla _{{\mathbf{V}}_f^{\left( i \right)}}}f 
	&= 4{\mathbf{A}}\left( {{{\bar \Psi }_{\bar G}}} \right)\left( {{{\mathbf{A}}^H}\left( {{{\bar \Psi }_{\bar G}}} \right){\mathbf{V}}_f^{}{\mathbf{V}}_f^H{\mathbf{A}}\left( {{{\bar \Psi }_{\bar G}}} \right) - {\mathbf{I}}_{\bar G}} \right) \nonumber \\
	&\quad \times {{\mathbf{A}}^H}\left( {{{\bar \Psi }_{\bar G}}} \right){\mathbf{V}}_f^{} \nonumber \\
	&= 4{\mathbf{A}}\left( {{{\bar \Psi }_{\bar G}}} \right){{\mathbf{A}}^H}\left( {{{\bar \Psi }_{\bar G}}} \right){\mathbf{V}}_f^{} \nonumber \\
	&\quad \times \left( {{\mathbf{V}}_f^H{\mathbf{A}}\left( {{{\bar \Psi }_{\bar G}}} \right){{\mathbf{A}}^H}\left( {{{\bar \Psi }_{\bar G}}} \right){\mathbf{V}}_f - {\mathbf{I}}_{N_R}} \right). \label{deqn_ex49}
\end{align}
		After obtaining the Riemannian gradient $gra{d_{{\mathbf{V}}_f^{\left( i \right)}}}f$, we can use the conjugate gradient method to update the search direction ${\mathbf{V}}_f^{(i + 1)}$ in the manifold space iteratively. The entire algorithm is shown in Algorithm \ref{alg:RMO}.

	With the optimized ${\mathbf{V}}_f^{}$, we minimize the total coherence of the measurement matrix $\bar {\bm{{\Gamma}}} $, which leads to improved estimation accuracy when using CS algorithms to solve the problem in (\ref{deqn_ex37}), such as the OMP algorithm. And the result ${{\bm{\bar \beta }}_{\bar G}}$ of (\ref{deqn_ex37}) provides the channel estimation $\hat{\mathbf{h}} = {\mathbf{A}}\left( {{{{{\bar \Psi }}}_{\bar G}}} \right){{\bm{\bar \beta }}_{\bar G}}$. The whole TSR channel estimation algorithm is summarized as Algorithm~\ref{alg:TSR}.
	
	{\emph {Remark 2}:}
	According to (\ref{deqn_ex14}) and (\ref{deqn_ex15}), it is noted that there are some conditions for $\bar {\bm{{\Gamma}}}$ to hold if we expect to recover ${{{\bm{\bar \beta }}}_{\bar G}}$ accurately. For $d= \lambda /2$, $Q$ is equal to ${N_E}$ and ${\bar G}={N_E}K$.
	Due to (\ref{deqn_ex15}), the lower bound of $	\mu \left( {\bar{\bm{{\Gamma}}}}  \right)$ is given by 
	\begin{equation}
		\label{deqn_ex50_a}
		\mu \left( {\bar{\bm{{\Gamma }}}}  \right) \ge \sqrt {\frac{{{{N_E}K} - {N_R}}}{{{N_R}\left( {{{N_E}K} - 1} \right)}}} .
	\end{equation}
	One necessary condition for (\ref{deqn_ex50_a}) is ${{N_E}K}>{N_R}$. If it is not satisfied, $\bar {\bm{{\Gamma}}}$ is a full column-rank matrix, which means we can get ${{{\bm{\bar \beta }}}_{\bar G}}$ by left multiplying the pseudo-inverse matrix of $\bar {\bm{{\Gamma}}}$ in (\ref{deqn_ex35}). For ${{N_E}K}>{N_R}$, based on the analysis in \cite {48}, another necessary condition to guarantee the recovery of ${{{\bm{\bar \beta }}}_{\bar G}}$ is 
	\begin{equation}
		\label{deqn_ex50_b}
		\frac{1}{{2K - 1}} > \sqrt {\frac{{{N_E}K - {N_R}}}{{{N_R}\left( {{N_E}K - 1} \right)}}}, 
	\end{equation}
	which transforms to the inequality
	\begin{equation}
		\label{deqn_ex50_c}
		{N_E}\left[ {{{\left( {2K - 1} \right)}^2} - {N_R}} \right] < 4{N_R}\left( {K - 1} \right).
	\end{equation}
	If ${{{\left( {2K - 1} \right)}^2} - {N_R}} \le 0$, there is no restriction on the number of ${N_E}$. On the contrary, if ${{{\left( {2K - 1} \right)}^2} - {N_R}}>0$, the number of ${N_E}$ should satisfy (\ref{deqn_ex50_c}) to guarantee unique recovery of ${{\bm{\bar \beta }}_{\bar G}}$.  
	
	\begin{algorithm}[t]
		\caption{TSR method for channel estimation}\label{alg:TSR}
		\begin{algorithmic}[1]
			\REQUIRE The number of signals $K$.
			\STATE Optimize ${\mathbf{v}}_r^{}$ with Algorithm \ref{alg:BFGS} according to \eqref{deqn_ex31_d}, and obtain ${\mathbf{V}}_r^{}$, under the assumption that ${\mathbf{v}}_r^{}={\mathbf{v}}_1^{}  =  \cdots  = {\mathbf{v}}_{{N_R}}$.
			\STATE Solve the optimization problem (\ref{deqn_ex27}) by DGMP in \cite{41} and obtain the set ${{\tilde \Theta }_{\tilde G}}$.
			\STATE  Obtain the candidate set of DoAs ${{\bar \Psi }_{\bar G}}$ using (\ref{deqn_ex24}).
			\STATE Design ${\mathbf{V}}_f^{}$ by the optimization problem (\ref{deqn_ex38}) with Algorithm \ref{alg:RMO}.
			\STATE Solve the optimization problem (\ref{deqn_ex37}) and obtain ${{{\bm{\bar \beta }}}_{\bar G}}$, and the estimated channel is $\hat{\mathbf{h}} = {\mathbf{A}}\left( {{{{{\bar \Psi }}}_{\bar G}}} \right){{\bm{\bar \beta }}_{\bar G}}$.
		\end{algorithmic}
	\end{algorithm}
	
	\section{Computational Complexity and Feasibility Analysis}
	\subsection{Computational Complexity}
The computational complexity of the proposed TSR method consists of the offline phase-shifting design and the online two-stage sparse recovery. 

\begin{table*}[t]
	\centering
		\caption{Computational Complexity of the Proposed TSR method}
		\label{tab:complexity}
		\begin{tabular}{>{\hspace{0pt}}m{0.32 \linewidth}>{\centering\arraybackslash\hspace{0pt}}m{0.58\linewidth}} 
			\toprule
			\textbf{Stage}                   & \textbf{Computational Complexity}  \\ 
			\midrule
			Offline Phase-Shifting Design    &$\mathcal{O}((M+I_r)N_E^4)$ \\
			Online Stage 1 (DGMP)                 & $\mathcal{O}(KN_R\tilde{G} + KR_cN_RJ_{\mathrm {loc}})$    \\
			Online Stage 2 (Manifold \& OMP)          & $\mathcal{O}(I_f(N\bar{G}+N_R\bar{G}^2) + KN_R\bar{G} + N_RK^2 + K^3)$ \\
			\textbf{Total Online Complexity} &$\mathcal{O}(KN_R(\tilde{G}+\bar{G}) + KR_cN_RJ_{\mathrm {loc}} + I_f(N\bar{G}+N_R\bar{G}^2) + N_RK^2 + K^3 )$ \\
			\bottomrule
		\end{tabular}
\end{table*}

First, the common phase-shifting vector $\mathbf v_r$ is optimized offline, since it only depends on the array geometry and the prescribed angular sector. In the L-BFGS implementation, constructing the matrix $\mathbf B=\sum_{m=1}^{M}\mathbf b_m\mathbf b_m^H$ requires $\mathcal O(MN_E^4)$ multiplications, while each gradient evaluation requires $\mathcal O(N_E^4)$ multiplications. Hence, with $I_r$ L-BFGS iterations, the offline complexity is $\mathcal O((M+I_r)N_E^4)$, which does not contribute to the real-time channel-estimation overhead.

In the first online stage, the identical phase responses across all RIMSAs fold the original angular dictionary of size $G$ into a reduced dictionary of size $\tilde G=G/Q$. Note that $Q=N_E$ for $d=\lambda/2$. Therefore, the dominant complexity of the DGMP-based principal phase estimation is $\mathcal O(KN_R\tilde G)$, with an additional local refinement cost $\mathcal O(KR_cN_RJ_{\mathrm {loc}})$, where $R_c$ is the number of refinement rounds and $J_{\mathrm loc}$ is the number of local grid points examined in each round. This is significantly lower than the $\mathcal O(KN_RG)$ complexity required by a direct search over the original angular grid.

In the second online stage, each estimated principal phase generates $Q = N_E$ candidate DoAs, leading to a candidate set of size $\bar G=KN_E$. The block-diagonal structure of the analog combining matrix $\mathbf V_f=\mathrm{blkdiag}(\mathbf v_{f,1},\ldots,\mathbf v_{f,N_R})$ is exploited when forming the effective candidate measurement matrix, reducing the cost from that of a dense matrix product to $\mathcal O(N\bar G)$. The dominant cost in each manifold iteration comes from evaluating the
objective function and its gradient. By first computing $\mathbf B_f=\mathbf V_f^H\mathbf A(\bar\Psi_{\bar G})$ and then $\mathbf B_f^H\mathbf B_f$, this cost is $\mathcal O(N\bar G+N_R\bar G^2)$ per iteration, where lower-order operations such as retraction and stopping criterion checking are omitted. With $I_f$ manifold iterations, the corresponding complexity is $\mathcal O(I_f(N\bar G+N_R\bar G^2))$. Finally, the OMP recovery over the reduced candidate set requires $\mathcal O(KN_R\bar G+N_RK^2+K^3)$ multiplications, whose dominant term is $\mathcal O(KN_R\bar G)$. Since $\bar G=KN_E$, the second-stage sparse recovery is performed over a much smaller candidate set than the original grid.

To provide a more intuitive overview, the computational complexity of the proposed TSR method is summarized in Table \ref{tab:complexity}.

\subsection{Feasibility}
From a practical deployment perspective, the computational burden is significantly alleviated by the system architecture. First, the optimization of $\mathbf{v}_r$  relies solely on the array geometry and the sector coverage requirements, independent of CSI. Consequently, the optimized phase-shifting vector can be precomputed offline and stored for real-time use, incurring no online optimization latency. For the online phase, the proposed TSR method exhibits high feasibility. In the first stage, the specific RIMSA design reduces the effective dictionary size from $G$ to $\tilde{G} = G/N_E$, substantially lowering the search complexity compared to conventional grid-based CS methods that search over the full $G$. Although the second stage involves manifold optimization, it operates on a reduced parameter set $\bar\Psi_{\bar G}$ derived from the first stage. For parameters $N=64, N_R=8$ used in our simulations, the matrix dimensions $64 \times 8$ in Eq. (\ref{deqn_ex49}) are small, involving standard matrix-vector multiplications that are highly parallelizable on modern digital signal processors. Therefore, the proposed method achieves a favorable trade-off between estimation accuracy and runtime latency.

	\section{Simulation Results}
	To numerically evaluate the proposed method and demonstrate its 
	superiority over existing methods, we present simulation results in this section. In these simulations, the fading coefficients ${\beta}_k$ for the multiple propagation paths are set with equal power allocation. We adopt the normalized mean square error (NMSE) as the performance metric of the channel estimation, which is defined as
	\begin{align}
		NMSE = \frac{1}{N_\mathrm{MC}}\sum\limits_{n = 1}^{N_\mathrm{MC}} {\frac{{{{\left\| {{{\widehat {\mathbf{h}}}_n} - {{\mathbf{h}}_n}} \right\|}_2^2}}}{{{{\left\| {{{\mathbf{h}}_n}} \right\|}_2^2}}}},\label{deqn_ex51} 
	\end{align}
	where $N_\mathrm{MC}$ is the total number of Monte Carlo simulations, ${{\mathbf{h}}_n}$ represents the true channel value of the $n$-th trial, and $\hat {\mathbf h}_n$ denotes the estimated channel value. The signal-to-noise ratio (SNR) is defined as $P/{\sigma_z^2}$, where $P$ represents the transmit power of a single pilot symbol. The interest zone is set to $\left[-\pi/3 ,\pi/3 \right]$, and ${{\mathbf{v}}_r}$ is optimized offline using Eq. (\ref{deqn_ex30}).
	
		\begin{figure}[!t]
		\centering
		\includegraphics[width=0.85\linewidth]{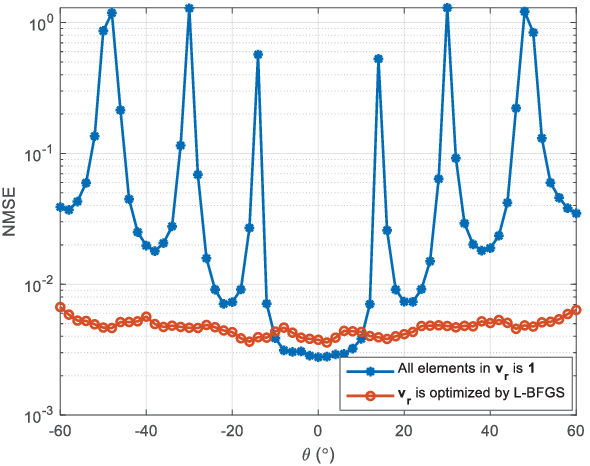}
		\caption{NMSE performance versus the direction of impinging signal for various ${{\mathbf{V}}_r}$.}
		\label{Fig.8}
	\end{figure}	
	The impact of metasurface antenna response on channel estimation performance is first investigated, and Fig. \ref{Fig.8} shows the NMSE of the single path channel estimation  over the direction of impinging signal, where $N = 64$, ${N_R} = 8$, and $SNR=0$ dB. We compare two scenarios: (1) using the identical phase-shifting vector ${{\mathbf{v}}_r}={\mathbf{1}}_{N_E}$, and (2) optimizing ${{\mathbf{v}}_r}$ via (\ref{deqn_ex30}), whose normalized RIMSA radiation patterns are illustrated in Fig. \ref{fig_5}(a) and  Fig. \ref{fig_5}(b), respectively. It can be observed that there are many peaks in the NMSE of the scenario that ${{\mathbf{v}}_r}={\mathbf{1}}_{N_E}$, primarily due to RIMSA pattern nulls at specific angles, which drastically reduces the received signal power. These results validate the theoretical analysis establishing a direct connection between metasurface antenna response and channel estimation accuracy. Specifically, larger antenna response amplitudes correspond to lower NMSE values, as the enhanced radiation efficiency increases the effective receive SNR of the RIMSA array.
		
	\begin{figure}[!t]
		\centering
		\includegraphics[width=0.91\linewidth]{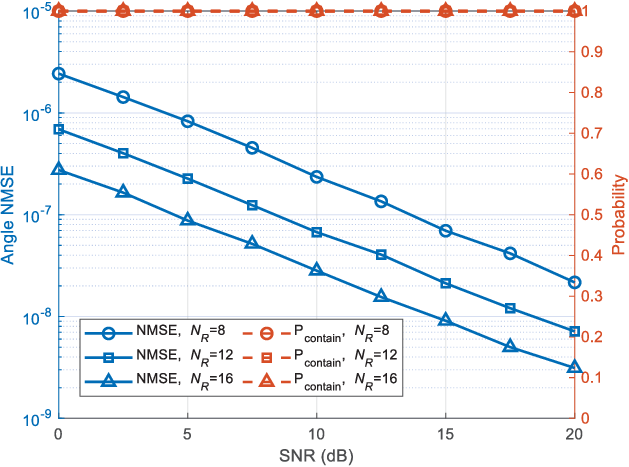}
		\caption{First-stage angle NMSE and containment probability versus SNR for different numbers of metasurface antennas $N_R$.}
		\label{Fig.9}
	\end{figure}
	\begin{figure}[!t]
		\centering
		\includegraphics[width=0.86\linewidth]{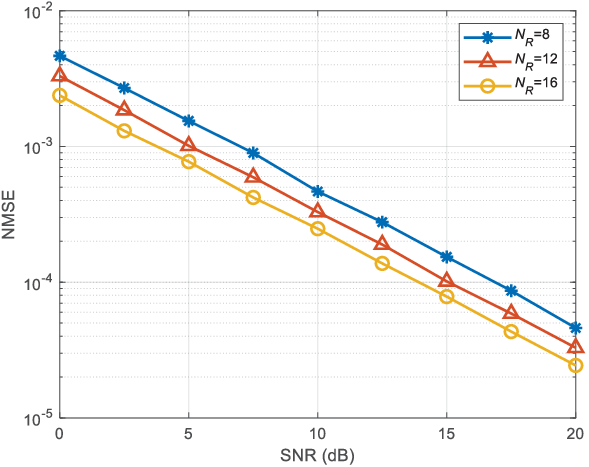}
		\caption{NMSE performance versus SNR for different numbers of metasurface antennas ${N_R}$.}
		\label{Fig.13}
	\end{figure}

	The sensitivity of the first stage is then analyzed using a single random path to validate the reliability of the proposed sequential estimation method. 	Since the TSR method relies on the candidate DoA set generated in the first stage, the first-stage design aims to retain sufficiently accurate candidates for all physical propagation paths.
	To comprehensively illustrate this mechanism, Fig. \ref{Fig.9} plots the angle NMSE (left axis) and the containment probability $P_\text{contain}$ (right axis) versus SNR for varying $N_R$ with $N_E=8$. Here, an ideal disambiguation is applied to evaluate the accuracy of the candidate angles. 
	Containment is considered successful if the closest candidate lies within half a grid interval of the true DoA in the direction-cosine domain, i.e.,	$\min_{\psi\in\bar{\Psi}_{\bar G}}|\sin\theta-\sin\psi|<{1}/{G}$.
	As shown on the left axis, the angle NMSE steadily decreases with higher $N_R$ and SNR, confirming the numerical precision of the initial candidate angles. Moreover, the right axis demonstrates that $P_\text{contain}$ remains consistently at $100\%$ in our Monte Carlo simulations over the evaluated SNR range.  
	This provides empirical evidence that the first-stage candidate set achieves a high containment probability for the physical propagation paths.
	This robustness stems from the reduced column dimension of the measurement matrix in the first stage, which leads to a lower mutual coherence compared to conventional CS-based approaches as derived in Eq. (\ref{deqn_ex28}). The reduced coherence significantly mitigates the basis mismatch effect and ensures a highly reliable candidate subspace for the subsequent refinement stage.
	
	\begin{figure}[!t]
		\centering
		\includegraphics[width=0.85\linewidth]{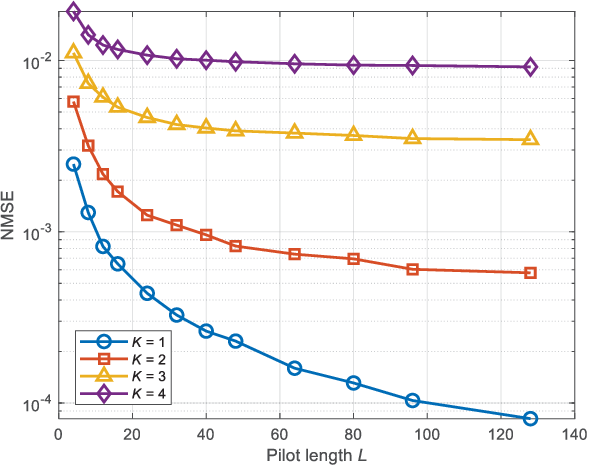}
		\caption{NMSE performance versus the pilot length $L$ for different numbers of propagation paths $K$.}
		\label{pilot_length}
	\end{figure}
	\begin{figure}[!t]
		\centering
		\includegraphics[width=0.85\linewidth]{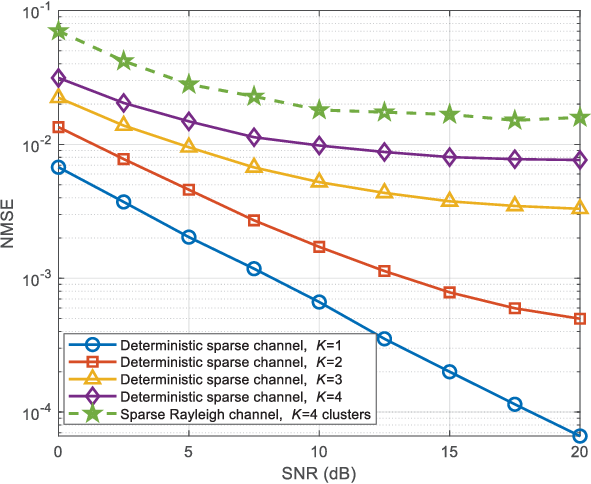}
		\caption{NMSE performance versus SNR for different numbers of propagation paths $K$.}
		\label{Fig_result_K}
	\end{figure}
	
	Based on the accurate candidate set provided by the first stage, Fig. \ref{Fig.13} further illustrates the NMSE versus SNR for different $N_R$ under the same simulation setup, and we can observe that the estimation accuracy improves progressively as $N_R$ increases. This trend is driven by the expanded array aperture enabled by additional RIMSA antennas, which enhances the effective observation dimensionality of received signals. Also, the reduced measurement matrix coherence design in the first stage provides a better starting point for the refinement in the second stage.

	The pilot overhead efficiency of the proposed TSR method is then explored. Fig. \ref{pilot_length} illustrates the NMSE performance versus the pilot length $L$ for different numbers of propagation paths $K$ with fixed $N_E=4$ and $N_R=16$. 
	The results show that the estimation accuracy consistently improves as $L$ increases. This improvement is mainly attributed to the enhanced effective observation quality obtained from a longer pilot sequence, which mitigates the noise effect through coherent averaging and thereby facilitates the subsequent sparse recovery. Conversely, for a fixed $L$, the NMSE degrades as $K$ increases. 
	This observation aligns with CS theory, since a larger $K$ makes the angular-domain channel less sparse, thereby increasing the difficulty of sparse recovery under the same pilot overhead.
	Notably, the proposed TSR method maintains satisfactory estimation accuracy even with a relatively small pilot length, demonstrating its robust capability to significantly reduce pilot overhead.

	\begin{figure}[!t]
		\centering
		\includegraphics[width=0.85\linewidth]{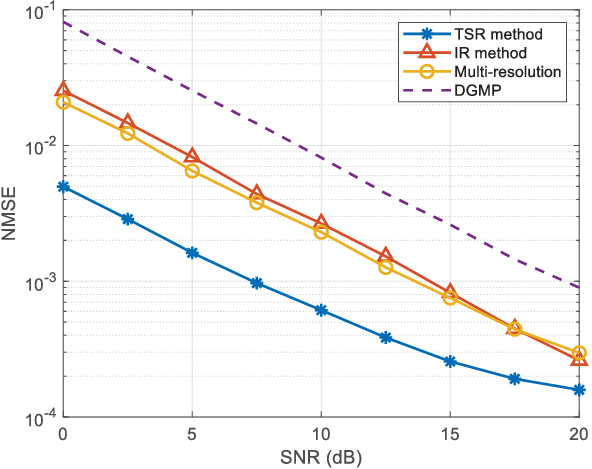}
		\caption{NMSE performance versus SNR for different hybrid MIMO methods.}
		\label{Fig.6}
	\end{figure}
	
	To further verify the robustness of the proposed TSR method against sparsity variations, Fig. \ref{Fig_result_K} compares the NMSE versus SNR for varying propagation paths $K$ with fixed $N_E=4$ and $N_R=16$. It can be observed that the estimation performance is best when $K=1$ and gradually degrades as $K$ increases. This aligns with CS theory, where reduced angular sparsity and potential path overlap complicate signal recovery.  However, the TSR method maintains acceptable estimation accuracy even in denser multipath scenarios, i.e., $K=3, 4$. Furthermore, to evaluate robustness in more realistic environments, we additionally consider a clustered channel model, where $K=4$ scattering clusters each contribute to a resolvable path \cite{44}. The path gains are modeled as independent Rayleigh fading variables, i.e., $\beta_k \sim \mathcal{CN}(0, 1/K)$, to maintain a normalized average channel power. As illustrated in Fig.~\ref{Fig_result_K}, despite the additional randomness introduced by Rayleigh fading, the proposed method still achieves high estimation accuracy at high SNR, demonstrating its practical reliability in complex multipath scenarios.

	Figure \ref{Fig.6} presents the performance of different channel estimation methods with a hybrid MIMO architecture. The RIMSA array is equipped with 8 metasurface antennas, each RIMSA containing 8 elements, for a total of 64 metamaterial elements, and the simulated multipath channel assumes $K=2$. The optimization problem (\ref{deqn_ex27}) is addressed using the DGMP method, while the OMP algorithm is employed to solve (\ref{deqn_ex37}). For comparison, we show the estimation performance by solving the optimization problem  (\ref{deqn_ex12}) directly with DGMP. Moreover, to ensure relative fairness, we also use the manifold algorithm to optimize ${\mathbf{V}}$ in the measurement matrix ${\bm{{\Gamma}}}  = {\mathbf{V}}^H{\mathbf{A}}\left( {{{{\Psi }}_G}} \right)$ in (\ref{deqn_ex12}), which can guarantee that the solution of problem (\ref{deqn_ex12}) is more accurate. Additionally, we also show the estimation performance of IR method in \cite{43} and multi-resolution method in \cite{44}. As evidenced by the results, the TSR method outperforms these methods, attributed to its minimized measurement matrix coherence, which significantly improves estimation performance.	
	\begin{figure}[!t]
		\centering
		\includegraphics[width=0.84\linewidth]{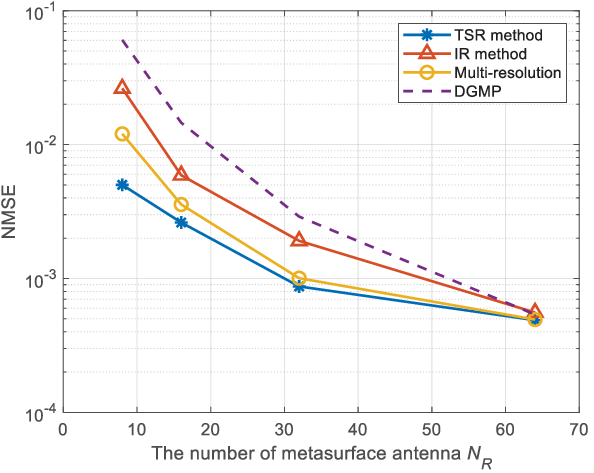}
		\caption{NMSE performance versus the number of RF chains with $N = 64$.}
		\label{Fig.12}
	\end{figure}
	
	Furthermore, Fig. \ref{Fig.12} illustrates the channel estimation performance as a function of the number of metasurface antennas under a fixed total number of metamaterial elements $N = 64$. For the multipath channel model with $K=2$ and $SNR=0$~dB, the estimation accuracy improves progressively as the number of metasurface antennas increases. That is because the dimension of observation sequences increases with the number of metasurface antennas. The results demonstrate that the proposed two-stage method achieves significantly higher estimation precision than direct channel estimation techniques when the number of metasurface antennas is fewer than the total metamaterial elements. This performance gain is primarily attributed to the specific design of the first stage, where the column dimension of the measurement matrix is reduced from $G$ to $\tilde{G} = G/Q$. Also, all curves eventually converge to the same minimum NMSE point. This is because when $N_R = N$, the hybrid analog-digital receiver architecture evolves into a fully digital structure, where the direct access to all spatial measurements eliminates the performance gaps among different hybrid MIMO methods.

	\section{Conclusion}
	In this paper, a two-stage refinement channel estimation method is proposed based on CS theory. 
	In the first stage, identical phase-shifting vectors are applied across all RIMSAs to reduce the effective dictionary dimension and the mutual coherence of the measurement matrix. Meanwhile, the common phase-shifting vector is optimized to avoid pattern nulls over the sector of interest, and then a candidate DoA set with DoA ambiguity is obtained. 
	In the second stage, the phase response of every metamaterial element is further optimized to resolve the DoA ambiguity and then refine the DoAs and channel coefficient estimates.
	Simulation results confirm that the proposed TSR method significantly outperforms existing methods in estimation accuracy. Moreover, our results validate the method's robustness in eliminating pattern nulls via phase optimization and maintaining high accuracy in hardware-constrained and multipath scenarios, demonstrating its potential for efficient massive MIMO systems.
	For future work, we plan to investigate the impact of mutual coupling among closely spaced metasurface elements in practical implementations. Additionally, developing adaptive sparse recovery methods that do not rely on prior knowledge of the channel sparsity level remains another crucial research direction.


\begin{thebibliography}{99}
	
		\bibitem{1}
	F. Boccardi, R. W. Heath Jr., A. Lozano, T. L. Marzetta, and P. Popovski,
	``Five disruptive technology directions for 5G,'' \emph{IEEE Commun. Mag.},
	vol. 52, no. 2, pp. 74–80, Feb. 2014. 
	
	\bibitem{2}
	E. D. Carvalho, A. Ali, A. Amiri, M. Angjelichinoski, and
	R. W. Heath Jr., ``Non-stationarities in extra-large-scale massive
	MIMO,'' \emph{IEEE Wireless Commun.}, vol. 27, no. 4, pp. 74–80,
	Aug. 2020.
	
	\bibitem{3}
	J. Mo, A. Alkhateeb, S. Abu-Surra, and R. W. Heath, ``Hybrid
	architectures with few-bit ADC receivers: Achievable rates and energy-rate tradeoffs,'' \emph{IEEE Trans. Wireless Commun.}, vol. 16, no. 4,
	pp. 2274–2287, Apr. 2017.
	
	\bibitem{4}
	Q. Wu and R. Zhang, ``Towards smart and reconfigurable environment:
	Intelligent reflecting surface aided wireless network,'' \emph{IEEE Commun.
		Mag.}, vol. 58, no. 1, pp. 106–112, Jan. 2020.
	
	\bibitem{5}
	M. Di Renzo et al., ``Smart radio environments empowered by reconfigurable intelligent surfaces: How it works, state of research, and the road
	ahead,'' \emph{IEEE J. Sel. Areas Commun.}, vol. 38, no. 11, pp. 2450–2525,
	Nov. 2020.
	
	\bibitem{6}
	Q. Wu and R. Zhang, ``Intelligent reflecting surface enhanced wireless
	network via joint active and passive beamforming,'' \emph{IEEE Trans. Wireless
		Commun.}, vol. 18, no. 11, pp. 5394–5409, Nov. 2019.
	
	\bibitem{7}
	Q. Wu and R. Zhang, ``Joint active and passive beamforming
	optimization for intelligent reflecting surface assisted SWIPT under QoS
	constraints,'' \emph{IEEE J. Sel. Areas Commun.}, vol. 38, no. 8, pp. 1735–1748,
	Aug. 2020.
	
	\bibitem{8}
	B. Zheng, C. You, W. Mei, and R. Zhang, ``A survey on channel
	estimation and practical passive beamforming design for intelligent
	reflecting surface aided wireless communications,'' \emph{IEEE Commun.
		Surveys Tuts.}, vol. 24, no. 2, pp. 1035–1071, 2nd Quart., 2022.
	
	\bibitem{9}
	S. Zhang and R. Zhang, ``Capacity characterization for intelligent
	reflecting surface aided MIMO communication,'' \emph{IEEE J. Sel. Areas
		Commun.}, vol. 38, no. 8, pp. 1823–1838, Aug. 2020.
	
	\bibitem{10}
	T. Van Chien, L. T. Tu, S. Chatzinotas, and B. Ottersten, ``Coverage
	probability and ergodic capacity of intelligent reflecting surface enhanced communication systems,'' \emph{IEEE Commun. Lett.}, vol. 25, no. 1,
	pp. 69–73, Jan. 2021.
	
	\bibitem{11}
	G. Zhou, C. Pan, H. Ren, K. Wang, M. Di Renzo, and A. Nallanathan,
	``Robust beamforming design for intelligent reflecting surface aided MISO
	communication systems,'' \emph{IEEE Wireless Commun. Lett.}, vol. 9, no. 10,
	pp. 1658–1662, Oct. 2020.
	
	\bibitem{12}
	C. Huang, A. Zappone, G. C. Alexandropoulos, M. Debbah, and
	C. Yuen, ``Reconfigurable intelligent surfaces for energy efficiency
	in wireless communication,'' \emph{IEEE Trans. Wireless Commun.}, vol. 18,
	no. 8, pp. 4157–4170, Aug. 2019.
	
	
	
	\bibitem{13}
	J. Liu and H. Zhang, ``Height-fixed UAV enabled energy-efficient data
	collection in RIS-aided wireless sensor networks,'' \emph{IEEE Trans. Wireless
		Commun.}, vol. 22, no. 11, pp. 7452–7463, Nov. 2023.
	
	\bibitem{121}
	L. Dong and H.-M. Wang, ``Secure MIMO transmission via intelligent
	reflecting surface,'' \emph{IEEE Wireless Commun. Lett.}, vol. 9, no. 6,
	pp. 787–790, June 2020.
	
	\bibitem{122}
	L. Dong and H.-M. Wang, ``Enhancing secure MIMO transmission via
	intelligent reflecting surface,'' \emph{IEEE Trans. Wireless Commun.}, vol. 19,
	no. 11, pp. 7543–7556, Nov. 2020.
	
	\bibitem{15}
	W. Wang, W. Ni, and H. Tian, ``Multi-functional RIS-aided wireless communications,'' \emph{IEEE Internet Things J.}, vol. 10, no. 23,
	pp. 21133–21134, Dec. 2023.
	
	
	\bibitem{123}
	L. Dong, H. -M. Wang, and J. Bai, ``Active reconfigurable intelligent surface aided secure transmission,'' \emph{IEEE Trans. Veh. Technol.}, vol. 71, no. 2, pp. 2181–2186, Feb. 2022.
	
	\bibitem{14}
	R. Long, Y.-C. Liang, Y. Pei, and E. G. Larsson, ``Active reconfigurable
	intelligent surface-aided wireless communications,'' \emph{IEEE Trans. Wireless Commun.}, vol. 20, no. 8, pp. 4962–4975, Aug. 2021.
	
	
	\bibitem{16}
	F. Shu et al., ``Three high-rate beamforming methods for active IRS-aided wireless network,'' \emph{IEEE Trans. Veh. Technol.}, vol. 72, no. 11,
	pp. 15052–15056, Nov. 2023.
	

	\bibitem{19}
	N. Shlezinger, O. Dicker, Y. C. Eldar, I. Yoo, M. F. Imani, and
	D. R. Smith, ``Dynamic metasurface antennas for uplink massive MIMO
	systems,'' \emph{IEEE Trans. Commun.}, vol. 67, no. 10, pp. 6829–6843,
	Oct. 2019.
	
	\bibitem{20}
	H. Wang et al., ``Dynamic metasurface antennas based downlink massive
	MIMO systems,'' in \emph{Proc. IEEE 20th Int. Workshop Signal Process. Adv.
		Wireless Commun. (SPAWC)}, Cannes, France, July 2019, pp. 1–5.
	
	\bibitem{18}
	R. Deng et al., ``Reconfigurable holographic surfaces for
	future wireless communications,'' \emph{IEEE Wireless Commun.},
	vol. 28, no. 6, pp. 126–131, Dec. 2021.
	
	\bibitem{WangZhang2026}
	H.-M. Wang, S. Zhang, and L. Wang, ``RIMSA: Reconfigurable intelligent metasurface antenna for 6G communications and sensing,'' \emph{IEEE Wireless Commun.}, 2026, early access, doi: 10.1109/MWC.2026.3690101.
	
	\bibitem{HuangWang2026}
	Y. Huang, H. -M. Wang, Q. Yan and Z. Wang, ``LLM-RIMSA: Large language models driven reconfigurable intelligent metasurface antenna systems,'' \emph{IEEE J. Sel. Areas Commun.}, vol. 44, pp. 2479-2493, 2026.
	
	\bibitem{add4}
	X. Wei and H.-M. Wang, ``Multi-user downlink with reconfigurable intelligent metasurface antennas (RIMSA) array,'' \emph{IEEE Trans. Veh. Technol.}, vol. 74, no. 12, pp. 18914-18929, Dec. 2025.
	
	\bibitem{add3}
	J. Bai, H. -M. Wang and L. Jin, ``Dynamic agile reconfigurable intelligent surface antenna (DARISA) MIMO: DoF analysis and effective DoF optimization,'' \emph{IEEE Trans. Wireless Commun.}, vol. 25, pp. 2197-2212, 2026.
	
	
	
	\bibitem{52}
	X. Yu, J.-C. Shen, J. Zhang, and K. B. Letaief, ``Alternating minimization
	algorithms for hybrid precoding in millimeter wave MIMO systems,''
	\emph{IEEE J. Sel. Topics Signal Process.}, vol. 10, no. 3, pp. 485–500,
	Apr. 2016.
	
	\bibitem{HamidrezaKhaleghi2025}
	H. Khaleghi and A. Haskou, ``Optimized channel estimation strategies for RIS-aided communication,'' \emph{IEEE Commun. Lett.}, vol. 29, no. 3, pp. 453-456, Mar. 2025.
	
	\bibitem{DavidWilliam2024}
	D. William Marques Guerra, T. Abrão and E. Hossain, ``Channel estimation in RIS-aided mmWave wireless systems using matching pursuit with phase rotation,'' \emph{IEEE Trans. Wireless Commun.}, vol. 23, no. 10, pp. 13187-13201, Oct. 2024.
	
	\bibitem{29}
	O. T. Demir, E. Bjornson, and L. Sanguinetti, ``Channel modeling and
	channel estimation for holographic massive MIMO with planar arrays,''
	\emph{IEEE Wireless Commun. Lett.}, vol. 11, no. 5, pp. 997–1001, May 2022.
	
	\bibitem{AnzhengTang2026}
	A. Tang, S. Song, C. -Y. Tsui, R. C. de Lamare and M. Debbah, ``Channel estimation for holographic MIMO systems with mutual coupling awareness,'' in \emph{Proc. IEEE Int. Conf. Acoust., Speech Signal Process. (ICASSP)}, Barcelona, Spain, 2026, pp. 21952-21956.
	
	
	\bibitem{35}
	M. Rezvani and R. Adve, ``Channel estimation for dynamic metasurface antennas,'' \emph{IEEE Trans. Wireless Commun.}, vol. 23, no. 6,
	pp. 5832–5846, June 2024.
	
	\bibitem{RuoyuZhang2025}
	R. Zhang et al., ``Tensor-based channel estimation for extremely large-scale MIMO-OFDM with dynamic metasurface antennas,'' \emph{IEEE Trans. Wireless Commun.}, vol. 24, no. 7, pp. 6052-6068, July 2025.
	
	\bibitem{add1}
	M. Ghermezcheshmeh and N. Zlatanov, ``Parametric channel estimation for LoS dominated holographic massive MIMO systems,'' \emph{IEEE Access}, vol. 11, pp. 44711–44724, 2023.
	
	\bibitem{add2}
	S. Chen, B. Sima, F. Xi, W. Wu, and Z. Liu, ``Super-resolution DOA estimation using dynamic metasurface antenna,'' in \emph{Proc. 14th Eur. Conf. Antennas Propag. (EuCAP)}, Copenhagen, Denmark, Mar. 2020, pp. 1–4.
	
	
	\bibitem{39}
	J. Lee, G.-T. Gil, and Y. H. Lee, ``Channel estimation via
	orthogonal matching pursuit for hybrid MIMO systems in millimeter wave communications,'' \emph{IEEE Trans. Commun.}, vol. 64, no. 6,
	pp. 2370–2386, June 2016.
	
	\bibitem{40}
	X. Ge et al., ``Training beam design for channel estimation in hybrid
	mmWave MIMO systems,'' \emph{IEEE Trans. Wireless Commun.}, vol. 21, no. 9,
	pp. 7121–7134, Sept. 2022.
	
	\bibitem{JavadMirzaei2021}
	J. Mirzaei, S. ShahbazPanahi, F. Sohrabi and R. Adve, ``Hybrid analog and digital beamforming design for channel estimation in correlated massive MIMO systems,'' \emph{IEEE Trans. Signal Process.}, vol. 69, pp. 5784-5800, 2021.
	
	\bibitem{41}
	Z. Gao et al., ``Channel estimation for millimeter-wave massive MIMO
	with hybrid precoding over frequency-selective fading channels,'' \emph{IEEE
		Commun. Lett.}, vol. 20, no. 6, pp. 1259–1262, June 2016.
	
	\bibitem{42}
	W. Chen, Y. Han, S. Jin, and H. Sun, ``Efficient multiband channel
	reconstruction and tracking for hybrid mmWave MIMO systems,'' \emph{IEEE
		Trans. Commun.}, vol. 69, no. 12, pp. 8501–8517, Dec. 2021.
	
	\bibitem{BiqingQi2019}
	B. Qi, W. Wang and B. Wang, ``Off-grid compressive channel estimation for mm-Wave massive MIMO with hybrid precoding,''  \emph{IEEE Commun. Lett.}, vol. 23, no. 1, pp. 108-111, Jan. 2019.
	
	\bibitem{42a}
	J. Rodríguez-Fernández, N. González-Prelcic, K. Venugopal, and 
	R. W. Heath, ``Frequency-domain compressive channel estimation for
	frequency-selective hybrid millimeter wave MIMO systems,'' \emph{IEEE
		Trans. Wireless Commun.}, vol. 17, no. 5, pp. 2946–2960, May 2018.
	
	
	\bibitem{43}
	C. Hu, L. Dai, T. Mir, Z. Gao, and J. Fang, ``Super-resolution channel
	estimation for mmWave massive MIMO with hybrid precoding,'' \emph{IEEE
		Trans. Veh. Technol.}, vol. 67, no. 9, pp. 8954–8958, Sept. 2018.
	
	\bibitem{44}
	A. Alkhateeb, O. El Ayach, G. Leus, and R. W. Heath, ``Channel
	estimation and hybrid precoding for millimeter wave cellular systems,''
	\emph{IEEE J. Sel. Topics Signal Process.}, vol. 8, no. 5, pp. 831–846, Oct.
	2014.

	
	
	\bibitem{47}
	L. Zelnik-Manor, K. Rosenblum, and Y. C. Eldar, ``Sensing matrix
	optimization for block-sparse decoding,'' \emph{IEEE Trans. Signal Process.},
	vol. 59, no. 9, pp. 4300–4312, Sept. 2011.
	
	\bibitem{48}
	L. R. Welch, ``Lower bounds on the maximum cross correlation
	of signals (Corresp.),'' \emph{IEEE Trans. Inf. Theory}, vol. IT-20, no. 3,
	pp. 397–399, May 1974.
	
	\bibitem{FoucartRauhut2013}
	S. Foucart and H. Rauhut, \emph{A Mathematical Introduction to Compressive Sensing}. New York, NY, USA: Birkh{\"a}user, 2013.
	
	
	\bibitem{49}
	E. J. Candes and T. Tao, ``Decoding by linear programming,'' \emph{IEEE Trans.
		Inf. Theory}, vol. 51, no. 12, pp. 4203–4215, Dec. 2005.
	
	\bibitem{50}
	J. M. Duarte-Carvajalino and G. Sapiro, ``Learning to sense sparse
	signals: Simultaneous sensing matrix and sparsifying dictionary
	optimization,'' \emph{IEEE Trans. Image Process.}, vol. 18, no. 7, pp. 1395–1408, Jul.
	2009.
	
	\bibitem{51}
	J. Tropp, ``Greed is good: Algorithmic results for sparse approximation,'' \emph{IEEE Trans. Inf. Theory}, vol. 50, no. 10, pp. 2231–2242,
	Oct. 2004.
	
	
	\bibitem{54}
	H.-M. Wang, J. Bai, and L. Dong, ``Intelligent reflecting surfaces assisted
	secure transmission without eavesdropper's CSI,'' \emph{IEEE Signal Process.
		Lett.}, vol. 27, pp. 1300–1304, 2020.
	
	
	\bibitem{60}
	Y. Wang, X. Wang, H. Liu, and Z. Luo, ``On the design of constant modulus
	probing signals for MIMO radar,'' \emph{IEEE Trans. Signal Process.}, vol. 60,
	no. 8, pp. 4432–4438, Aug. 2012.
	
	
	\bibitem{57}
	J. Nocedal and S. J. Wright, ``Quasi-Newton Methods,'' in \emph{Numerical Optimization}, 2nd ed. New York, NY, USA: Springer, 2006, pp. 135--163.
	
	
	
	
\end{thebibliography}

\end{document}